\documentclass[%
 aip,
 amsmath,amssymb,
 reprint,%
floatfix]{revtex4-1}

\usepackage{graphicx}
\usepackage{dcolumn}
\usepackage{dirtytalk}
\usepackage{bm}

\usepackage[utf8]{inputenc}
\usepackage[T1]{fontenc}
\usepackage{mathptmx}
\usepackage{amsmath}
\usepackage{xcolor}
\usepackage{float}
\usepackage{hyperref}

\setcitestyle{super}

\newcommand*{\citen}[1]{%
  \begingroup
    \romannumeral-`\x 
    \setcitestyle{numbers}%
    \cite{#1}%
  \endgroup   
}

\newenvironment{Notes}
{\begin{quote}\small\tt Note to Dhiman: \ }
{\end{quote}}

\newenvironment{Notes_2}
{\begin{quote}\small\tt Note to Ioan: \ }
{\end{quote}}

\newenvironment{Notes_3}
{\begin{quote}\small\tt Note to DL Mobley: \ }
{\end{quote}}

\newcommand{\bno}{\begin{Notes}}
\newcommand{\eno}{\end{Notes}\noindent}

\begin{document}

\preprint{AIP/123-QED}

\title[]{Kinetics and Free Energy of Ligand Dissociation Using Weighted Ensemble Milestoning}

\author{Dhiman Ray}
\affiliation{ 
Department of Chemistry, University of California Irvine, California 92697, USA
}
\author{Trevor Gokey}
\affiliation{ 
Department of Chemistry, University of California Irvine, California 92697, USA
}
\author{David L. Mobley}
  \affiliation{ 
Department of Chemistry, University of California Irvine, California 92697, USA
}
 \affiliation{Department of Pharmaceutical Sciences, University of California Irvine, California 92697, USA}

\author{Ioan Andricioaei}%
 \email{andricio@uci.edu.}
\affiliation{ 
Department of Chemistry, University of California Irvine, California 92697, USA
}
\affiliation{ 
Department of Physics and Astronomy, University of California Irvine, California 92697, USA
}

\begin{abstract}
We consider the recently developed weighted ensemble milestoning (WEM) scheme [J. Chem. Phys. 152, 234114 (2020)], and test its capability of simulating ligand-receptor dissociation dynamics. We performed WEM simulations on the following host-guest systems: Na+/Cl- ion pair and 4-hydroxy-2-butanone (BUT) ligand with FK506 binding protein (FKBP). As proof or principle, we show that the WEM formalism reproduces the Na+/Cl- ion pair dissociation timescale and the free energy profile obtained from long conventional MD simulation. To increase accuracy of WEM calculations applied to kinetics and thermodynamics in protein-ligand binding, we introduced a modified WEM scheme called weighted ensemble milestoning with restraint release (WEM-RR), which can increase the number of starting points per milestone without adding additional computational cost. WEM-RR calculations obtained a ligand residence time and binding free energy in agreement with experimental and previous computational results. Moreover, using the milestoning framework, the binding time and rate constants, dissociation constant and the committor probabilities could also be calculated at a low computational cost. We also present an analytical approach for estimating the association rate constant ($k_{\text{on}}$) when binding is primarily diffusion driven. We show that the WEM method can efficiently calculate multiple experimental observables describing  ligand-receptor binding/unbinding and is a promising candidate for computer-aided inhibitor design.
\end{abstract}

\maketitle

\section{Introduction}
\noindent
Protein-ligand interactions play crucial roles in modulating the biological processes inside the cell. Understanding the molecular details of ligand binding and unbinding is necessary not only to gain fundamental insight into molecular recognition, but also to facilitate rational design of drugs and inhibitors, thereby tuning the functionality of a particular protein in a desired way \cite{Hajduk2007ALearned,Pramanik2019CanDissociation}. Molecular dynamics (MD) simulation, with its use of physics-based models to represent the interatomic forces and propagate the dynamics in time, has become a centerpiece method for studying the dynamics of biomolecules in atomistic detail\cite{McCammon1977DynamicsProteins}. Pioneering work -- much of it done with the CHARMM simulation package\cite{Brooks1983CHARMM:Calculations} celebrated in this journal issue -- has spawned applications that revealed the role of atomic motions in conformational switching, folding and ligand binding in proteins and other bio-molecules \cite{Karplus2002MolecularBiomolecules}.

Atomistic MD simulation has been shown to capture well the physics of ligand-receptor binding and has been successfully used to calculate the binding free energies \cite{Simonson2002FreeRecognition}, residence times \cite{Tiwary2015KineticsSteps} and binding and unbinding pathways \cite{Shan2011HowSite} for a variety of systems. 

However, the majority of the biologically interesting protein-ligand systems are characterized by ligand residence times that span long intervals, from a few milliseconds to multiple hours. 
Such binding or unbinding processes fall in the realm of rare events \cite{Zwier2010,Chong2017}. Rare events are particularly challenging from a molecular simulation point of view for primarily two reasons. First, the timescales involved are often beyond the capacity of currently available computer power, which can at best simulate up to multiple microseconds for protein systems. Second, because of the long waiting time before a transition is to occur, most of the invested computational power is wasted in the free energy (local) minimum corresponding to the initial state. 
Simulations of the binding process also encounter the challenge of needing to find one (or several) entry path(s) via which the ligand diffuses into the binding pocket; this necessitates additional significant investment of computation effort \cite{Chong2017,Ray2020WeightedSimulations,Bagchi2018StatisticalScience}. 

To overcome these difficulties, effort has been expanded on the development of enhanced sampling methods that apply artificial biases to accelerate events of interest. 
Methods like umbrella sampling (US) \cite{Torrie1977,Roux1995TheSimulations}, steered molecular dynamics (SMD) \cite{Park2004CalculatingSimulations}, metadynamics (MtD) \cite{Laio2002}, adaptive biasing force (ABF) \cite{Darve2001,Comer2015}, and accelerated molecular dynamics (aMD) \cite{Hamelberg2004AcceleratedBiomolecules} could sample the conformational space of bio-molecules at significantly lower computational cost and consequently gained popularity and became widespread in computational biophysics. Although the free energy landscape along a chosen set of collective variables can be computed from these methods, recovering kinetic information is extremely difficult because the dynamics of the system is rendered artificial by the application of unnatural biases. As such, the biases distort the unbinding mechanism to a significant degree \cite{Dickson2016LigandMechanisms}.

An attempt to solve the kinetics problem using a master equation based approach, from unbiased simulation data, is commonly referred to as the Markov state modeling (MSM) \cite{Pande2010,Voelz2010MolecularNTL91-39} in the MD simulation literature.  

MSMs have experienced considerable success in ligand binding/unbinding simulations over the past two decades owing to the rapid development of fast computers and the consequent availability of a large amount of MD simulation data \cite{Held2011MechanismsMutations,Mondal2018}. However, this approach often falls short in the presence of very high free energy barriers, where a low number of transitions results in overestimation of the transition rates (\textit{over}-estimation because one misses the slower transitions that make up tail of the first passage time distribution; this is a particularly severe problem in MSM studies that use only a handful of transition events to infer timescales).  

The weighted ensemble (WE) method, introduced by Huber and Kim more than 20 years ago \cite{Huber1996}, is a pioneering example of a technique that tackled the problem of computing \emph{time-dependent} properties (e.g., rate constants, mean first -passage times, etc.) that, for complex dynamical system, could not be derived directly from free energy landscapes obtained via the otherwise enhanced sampling techniques\cite{Torrie1977,Roux1995TheSimulations} available at that time. 
Zhang et al. recently showed that WE is statistically exact, and applied it both MD and MC sampling to study a wide range of processes modulated by rare events\cite{Zhang2010b}. The implementation of this technique in user friendly packages like WESTPA has led to an extensive use of weighted ensemble based methods in biophysical problems \cite{Zwier2015,Zwier2016,Saglam2017FlexibilityAnalogues,Saglam2016HighlyModels,Saglam2019Protein-proteinSimulations}.

The binding and unbinding dynamics of many host guest systems have been studied using WE. The examples include association of the Na+/Cl- ions, K+ binding to 18-crown-6 ether \cite{Zwier2011}, protein-peptide binding \cite{Zwier2016,Saglam2017FlexibilityAnalogues}, and protein-protein binding \cite{Saglam2016HighlyModels,Saglam2019Protein-proteinSimulations}. Recently, Saglam and Chong used a steady state weighted ensemble methodology to calculate the rate constant of a multi-microsecond protein–protein association in explicit solvent \cite{Saglam2019Protein-proteinSimulations}, spending only <1\% of the simulation time required to construct a converged MSM for the same system\cite{Plattner2017CompleteModelling} from conventional MD simulations. 

Novel variants of the weighted ensemble method have been used to study multiple long timescale protein-ligand unbinding kinetics and pathways\cite{Dickson2016LigandMechanisms,Dickson2017MultipleWExplore,Donyapour2019REVO:Optimization,Lotz2018UnbiasedInteractions,Dixon2020Membrane-mediatedTSPO}. Donyapour and coworkers used WExplore \cite{Dickson2014WExplore:Algorithm} and resampling of ensembles by variation optimization (REVO) \cite{Donyapour2019REVO:Optimization} to calculate the unbinding time of benzamidine ligand from trypsin within an order of magnitude accuracy. In an impressive attempt, Lotz and Dickson used steady state WExplore scheme to simulate the very slow unbinding process of 1-trifluoromethoxyphenyl-3-(1-propionylpiperidin-4-yl)-urea, or TPPU from soluble epoxide hydrolase (sEH) from microsecond simulations. The calculated residence time (42s) was only a little over an order of magnitude away from the experimental result (11 min) \cite{Lotz2018UnbiasedInteractions}. Recently a multi-microsecond REVO simulation study of PK-11195 ligand dissociation from translocator protein (TSPO), a membrane bound receptor, has revealed a lipid assisted unbinding mechanism. The calculated residence times (4.1 min - 260 min) starting from different binding poses were within one order of magnitude from the experimental residence time of 34 min \cite{Dixon2020Membrane-mediatedTSPO}.

Developed independently about 16 years ago, the milestoning technique of Elber and coworkers \cite{Faradjian2004} is another powerful method that can be used to compute timescales, as well as free energy profiles along suitably chosen directions in configuration space. In a recent work, the mathematical details of milestoning theory have been clarified \cite{Bello-Rivas2015}, allowing the calculation of many thermodynamic and kinetic properties including mean first passage times \cite{West2007,Elber2007}, free energy profiles \cite{Ma2017TheTheory,Ma2018ProbingMilestoning,Cardenas2016}, committor probabilities \cite{Elber2017CalculatingMilestoning}, and time-correlation functions \cite{Grazioli2018}.

Recently, the milestoning method has been implemented in the SEEKR package, which led to a number of studies on the binding-unbinding kinetics of host-guest systems \cite{Votapka2017SEEKR,Jagger2018QuantitativeApproach,Ahn2020RankingApproach} involving a combination of molecular dynamics (MD) and Brownian dynamics (BD) simulation \cite{Votapka2015MultiscaleMilestoning}. A major limitation of the milestoning scheme is that the reaction coordinate and the milestone positions must be predefined. In a recent study, principal component based reaction coordinates were used in milestoning to determine millisecond scale residence time of protein-ligand system \cite{Tang2020TransientDesign}. For placing milestones on rugged free energy landscapes with multiple minima, i.e., when intuition for a good reaction coordinate is lacking, an approach has been developed for identifying best milestones using machine-learning 
\cite{Grazioli2017AutomatedLearning}. 

The milestones need to be placed sufficiently far apart so that the memory of the starting milestone is lost before the trajectory hits one of the adjacent milestones \cite{West2007}. However it requires considerable computational effort to converge the transition statistics for milestones placed too far from each other. This problem motivated our previous effort to accelerate milestoning trajectories using biasing forces \cite{Grazioli2018a}. The resulting approach, wind assisted re-weighted milestoning (WARM) \cite{Grazioli2018a}, helps climbing steep energy landscapes and trajectory reweighting recovers the correct, bias-free kinetics. However, when the free energy surface is flat, i.e., in the diffusive regime, the reweighted averages and standard deviations of observables are not significantly better than normal milestoning \cite{Wang2020MilestoningKinetics}.

As an alternative to improve sampling during milestoning, we suggested that weighted ensemble (WE) based methods can provide a great opportunity to enhance milestone-to-milestone transitions without applying artificial biasing forces\cite{Ray2020WeightedSimulations}. In our previous work we blended the two methods and developed a combined weighted ensemble milestoning (WEM) scheme that performs equilibrium WE simulations \cite{Suarez2014SimultaneousTrajectories} in-between milestones. WEM improves conventional milestoning by providing quicker convergence of the transition kernel, thereby allowing one to place the milestones sufficiently far from each other \cite{Ray2020WeightedSimulations}. The method brings in the mathematical framework of milestoning into the weighted ensemble framework for the calculation of free energy, kinetics and time correlation function from a master equation like approach. Moreover, it facilitates the parallelization of WE simulation over multiple milestones, significantly reducing the wall clock time for the simulation. Together, the WEM approach improves over both milestoning and weighted ensemble to compute many important experimental observables from short, easily obtainable MD simulations.

In this paper we apply our WEM scheme by addressing existing ligand-receptor models regularly studied in the biophysics MD community. As proof of principle, we first look into a simple model of the Na+/Cl- ion pair, which has previously been used as a primary test case for both WE \cite{Zwier2011} and milestoning \cite{Votapka2015MultiscaleMilestoning} 
Next we study the biologically relevant system of 4-hydroxy-2-butanone (BUT) ligand dissociation from FK506 binding protein (FKBP).

FKBP is present in a wide range of eukaryotic cells and functions as a protein folding chaperone for proline containing proteins \cite{Siekierka1989ACyclophilin}. FKBP was previously crystallized bound to BUT with 1.85 \AA$ $ resolution. The study also reported the dissociation constant ($K_D$) to be $500 \mu$M and the binding free energy ($\Delta G$) to be 18.9 kJ/mol, using fluorescence quenching measurement of the Trp59 residue near the binding pocket \cite{Burkhard2000X-rayEnergies}. The FKBP protein is one of the protein-ligand complexes which has been most widely studied with MD simulations \cite{Huang2011TheUnbinding,Dickson2016LigandMechanisms,Pan2017QuantitativeSimulations,Pramanik2019CanDissociation}. The residence time of BUT ligand is of the order of 10 ns, making exhaustive sampling of reversible binding and unbinding events possible through brute force MD simulation \cite{Pan2017QuantitativeSimulations}. Also there is no significant protein conformational change coupled to the ligand release pathway. This provides a great opportunity to study new simulation methods to test for their accuracy and efficiency in comparison to regular MD. 

\textcolor{black}{
An extensive computational study of the} dissociation of multiple ligands including BUT from FKBP protein has been performed by Huang and Caflisch \cite{Huang2011TheUnbinding} using the CHARMM22 force field \cite{MacKerell1998All-atomProteins} for the protein and the CHARMM general force field (CGenFF) \cite{Vanommeslaeghe2010CHARMMFields} for the BUT ligand. \textcolor{black}{Conformations sampled from multiple short MD runs were used to perform a complex network analysis from which the cut-based free energy profile (cut-based FEP) was computed following the method proposed by Krivov and Karplus \cite{Krivov2006One-dimensionalBarriers}.} The unbinding events were characterized by a distance of more than 15 \AA$ $ between the ligand center of mass and the FKBP binding site. The system was considered bound when that distance was less than 10 \AA. Huang and Caflisch could calculate the residence times and binding $\Delta G$ in agreement with the experiment \cite{Burkhard2000X-rayEnergies} along with the contribution of electrostatic and Van-der Waals interactions to the binding affinity \cite{Huang2011TheUnbinding}. The network analysis of the trajectory data also revealed multiple ligand poses of BUT inside the binding pocket and identified the residues, Asp37, Ile56 and Tyr82, as directly interacting with the ligand in its bound state \cite{Huang2011TheUnbinding}. 

Lotz and Dickson used their newly developed WExplore method \cite{Dickson2014WExplore:Algorithm} to perform steady state simulations to calculate the unbinding time and to elucidate the ligand release mechanism and pathways for BUT along with two other ligands \cite{Dickson2016LigandMechanisms}. They used the CHARMM36 force field \cite{Huang2017} for the protein, CGenFF force field \cite{Vanommeslaeghe2012AutomationCharges,Vanommeslaeghe2012AutomationTyping} for the ligand, and a reaction coordinate based on the root mean squared distance (RMSD) between the bound state of the ligand and the coordinates of the ligand during the simulation. Their results showed quantitative agreement with the work by Huang and Caflisch \cite{Huang2011TheUnbinding}. 

The most extensive work so far is by Pan et. al., who performed 30 $\mu$s of brute force MD simulation on the Anton supercomputer, starting from the FKBP-BUT crystal structure, to sample 277 reversible binding and unbinding events from which $\Delta G$, $K_D$, rate constants for binding ($k_{\text{on}}$) and unbinding ($k_{\text{off}}$) have been obtained \cite{Pan2017QuantitativeSimulations}. They also obtained identical results for binding free energy using free energy perturbation (FEP) method \cite{Wang2015AccurateField}. Their work was performed using  AMBER ff99SB*-ILDN force field  \cite{Hornak2006ComparisonParameters,Best2009OptimizedPolypeptides,Lindorff-Larsen2010ImprovedField} for protein and generalized AMBER force field (GAFF) \cite{Wang2004DevelopmentField} for the ligand. They used the distance between Trp59 residue and the ligand as the reaction coordinate and trajectories were considered unbound if the value of the distance is larger than 6 \AA. 

Pramanik et al. also performed multiple microseconds of conventional MD and metadynamics simulation for FKBP-BUT complex to understand the reliability of kinetic observables calculated from infrequent metadynamics approach \cite{Pramanik2019CanDissociation}. They used identical force field parameters as Pan et al. \cite{Pan2017QuantitativeSimulations} but a complex reaction coordinate obtained using the spectral gap optimization of order parameters (SGOOP) method. \cite{Tiwary2016HowUnbinding,Smith2018Multi-dimensionalFactorization,Tiwary2016SpectralSystems}. Their distance criterion for unbinding is as low as 1.8 \AA$ $ of separation between Trp59 and BUT \cite{Pramanik2019CanDissociation}. Their calculated unbinding time and free energy values are consistent with the work of Pan et al. but largely different from the experimental numbers \cite{Burkhard2000X-rayEnergies} and the extensive simulation results by Huang and Caflisch \cite{Huang2011TheUnbinding}.

The availability of both computational and experimental studies on the FKBP-BUT complex motivated us to choose this particular system for our current study. Here we aim to asses the accuracy and efficiency of the WEM method for simulating ligand-receptor binding-unbinding dynamics. 

The rest of our paper is organized as follows. Section \ref{sec:WEMscheme} provides a brief description of the WEM scheme and methods of error estimation followed by the computational details in section \ref{sec:methods}. In Section \ref{sec:results} we discuss our results for the Na+/Cl- and FKBP/BUT systems, and put them into context of previous studies. Lastly, in Section \ref{sec:conclusions} we conclude by mentioning the key advantages of WEM simulation along with possible directions of improvement and future research.

\section{Theory}
\subsection{Weighted ensemble milestoning (WEM) scheme}
\label{sec:WEMscheme}
In our previous paper we introduced the weighted ensemble milestoning (WEM) scheme and provided a detailed description of the algorithm \cite{Ray2020WeightedSimulations}. Here we only provide a brief overview. 

In the original milestoning method, high dimensional interfaces are placed along the reaction coordinate. They confer on configuration space a stratified structure in which multiple short MD trajectories are propagated in between the interfaces. The probability of transition between milestones is estimated by counting the fraction of trajectories that start from a given milestone and reach any of the nearest adjacent milestones. The lifetime of a milestone is defined as the average amount of time spent by trajectories staring from it before reaching nearby milestones. 

The transition probability between milestone $i$ and $j$ ($K_{ij}$) and the lifetime of milestone $i$ ($\overline{T}_i$) are given by $K_{ij} = \frac{N_{i \rightarrow j}}{N_i}$ and $\overline{T}_i = \frac{\sum\limits_{l=1}^N t_l}{N_i}$, where $N_i$ is the number of trajectories starting from milestone $i$, $N_{i \rightarrow j}$ is the number of such trajectories ending at milestone $j$, and $t_l$ is the time spent by the $l$th such trajectory before hitting either of the nearest milestones \cite{Bello-Rivas2015,West2007,Faradjian2004}. Along a one-dimensional reaction coordinate, which is the case considered in this paper, $j = i \pm 1$.

The gist of the WEM method is to further stratify the space between milestones by performing weighted ensemble simulations therein. We put an integer number of WE bins in between milestones. Trajectories are propagated starting from a given milestone. If the trajectories enter a new bin at the end of an iteration time $\delta t$, they are split into multiple trajectories or merged into one, maintaining an equal number of trajectories in each bin. The weights of the trajectories are redistributed during this splitting/merging process to conserve the total probability. A stochastic thermostat -- such as the Langevin thermostat used herein -- ensures that multiple trajectories generated from the same initial configuration follow different paths upon trajectory splitting (although subtleties regarding the choice of random seeds need to be considered \cite{Uberuaga2004SynchronizationExploitation}).

Apart from counting the number of trajectories hitting nearby milestones, we now also record their weights, as they have different probabilities. The lifetime of milestone $i$ in WEM is given by
\begin{equation}
\overline{T}_i = \sum_k t_k w_k,
\end{equation}
where the sum is over all trajectories hitting any of the adjacent milestones. The $t_k$ is the time spent by the $k$th trajectory before reaching a different milestone and $w_k$ is its weight. The elements of the transition kernel $\mathbf{K}$ are given by
\begin{equation}
\begin{split}
K_{ij} & = \sum_{k\in \Gamma(i\rightarrow j)} w_k; \quad j = i \pm 1 \\
& = 0;\quad \text{otherwise}
\end{split}
\label{eqn:kij_wem}
\end{equation}
where $\Gamma(i\rightarrow j)$ is the set of trajectories starting from milestone $i$ and reaching at $j$ before any other milestone. The $K_{ij}$ is zero for milestones which are not adjacent to each other because a trajectory cannot reach further milestones in 1D before reaching the nearest milestones.

The stationary flux through each milestone $q_{i}$ can be obtained from the elements of the left principal eigenvector of the transition kernel \cite{Bello-Rivas2015}:
\begin{equation}
\mathbf{q}^T \mathbf{K} = \mathbf{q}^T
\label{eqn:statflux}
\end{equation}
where superscript $T$ denotes transpose. Eq. \ref{eqn:statflux} is equivalent to solving the linear equations $\sum\limits_i q_{i} K_{ij} = q_{j}$. It may seem counter-intuitive that the principal eigenvector of $\mathbf{K}$ gives fluxes and not equilibrium probabilities as in case of a Markov process. A rational behind this is discussed in our previous work \cite{Ray2020WeightedSimulations}.

The equilibrium probability of milestone $i$, defined as the probability of a trajectory which has started from milestone $i$ and has not crossed any other milestones, can be obtained by:
\begin{equation}
P_{\text{eq},i} = q_{i} \overline{T}_i
\label{eqn:peq}
\end{equation}
and subsequently the free energy of milestone $i$ is computed as
\begin{equation}
\Delta G_i = -k_BT \ln \bigg(\frac{P_{\text{eq},i}}{P_{\text{eq},0}} \bigg)
\label{eqn:free-energy}
\end{equation}
where $P_{\text{eq},0}$ is a reference probability usually chosen to be that of the most probable milestone. 

Before the application of Equation \ref{eqn:statflux}, a reflective boundary condition has to be implemented in matrix $\mathbf{K}$ to avoid draining of the probabilities through the last milestone. For example, in a 3-milestone system the original transition kernel is modified in the following way \cite{Kirmizialtin2011RevisitingMilestoning,Ma2017FreeFlipping}:
\begin{equation}
    \mathbf{K} = 
    \begin{bmatrix} 
    0 & K_{12} & 0\\
    K_{21} & 0 & K_{23}\\
    0 & K_{32} & 0
    \end{bmatrix} \rightarrow 
    \begin{bmatrix} 
    0 & K_{12} & 0\\
    K_{21} & 0 & K_{23}\\
    0 & 1 & 0
    \end{bmatrix}
\end{equation}
For calculating the rate constant or mean first passage time a steady state flux has to be established. To achieve that the $\mathbf{K}$ matrix is modified in the following way (for a 3-milestone case) where the flux passing through the last milestone is fed back into the starting milestone \cite{Kirmizialtin2011RevisitingMilestoning,Ma2017FreeFlipping}:
\begin{equation}
    \mathbf{K} = 
    \begin{bmatrix} 
    0 & K_{12} & 0\\
    K_{21} & 0 & K_{23}\\
    0 & K_{32} & 0
    \end{bmatrix} \rightarrow 
    \begin{bmatrix} 
    0 & K_{12} & 0\\
    K_{21} & 0 & K_{23}\\
    1 & 0 & 0
    \end{bmatrix}
\end{equation}
After Equation \ref{eqn:statflux} is used to obtain the steady state flux, the mean first passage time can be obtained from it using the following expression \cite{Bello-Rivas2015,Ma2017FreeFlipping}:
\begin{equation}
    \langle \tau \rangle = \frac{\sum\limits_i q_{i} \overline{T}_i}{q_{f}}
    \label{eqn:MFPT}
\end{equation}
where $q_f$ denotes the steady state flux through the final or product milestone and the summation runs over all the milestones.

The mean first passage time of the reverse process can be obtained from a new transition kernel $\mathbf{K}'$ and lifetime vector $\mathbf{\overline{T}}'$, where $K'_{ij} = K_{M-i,M-j}$ and $\overline{T}'_i = \overline{T}_{M-i}$ for all $i$,$j$ 
where $M$ is the total number of milestones. Basically the first milestone is relabelled as the last milestone, the second one as the second last and so on.

So, for a protein ligand system, we can obtain both unbinding and binding times $\langle \tau \rangle_{\text{off}}$ and $\langle \tau \rangle_{\text{on}}$ using the milestoning framework. We can use them to obtain the unbinding and binding rate constants $k_{\text{off}}$ and $k_{\text{on}}$ using the following expression from Pan et. al. \cite{Pan2017QuantitativeSimulations} :
\begin{equation}
    \begin{split}
        &k_{\text{off}} = \frac{1}{\langle \tau \rangle_{\text{off}}} \\
        &k_{\text{on}} = \frac{1}{\langle \tau \rangle_{\text{on}}} v c^0 N_{av}
    \end{split}
    \label{eqn:koff-kon}
\end{equation}
where $v$ is the volume of the equilibrated simulation box, $c^0$ is the standard molar concentration i.e. 1.0 M, and $N_{av}$ is Avogadro's number. The dissociation constant $K_D$ is then calculated as
\begin{equation}
    K_D = \frac{k_{\text{off}}}{k_{\text{on}}}
    \label{eqn:KD}
\end{equation}
which gives a the standard binding free energy of
\begin{equation}
    \Delta G^{\circ} = -k_BT \ln K_D
    \label{eqn:delta-G}
\end{equation}
where we infer $K_D$ to be a unitless ratio relative to the standard concentration of 1 M \cite{Gilson1997TheReview}.

\textcolor{black}{Alternatively, the binding free energy can be obtained from the free energy profile or the potential of mean force (PMF) \cite{Souza2020ProteinligandModel}. For reaction coordinates based on radial distance (like the ligand-receptor distance ($r$) used in the current work) the PMF can be computed from the free energy profile $G(r)$ by correcting for the Jacobian factor:
\begin{equation}
    {\rm PMF}(r) = G(r) + 2 k_BT \ln(r)
    \label{eqn:jacobian}
\end{equation}
From the PMF the binding constant $K_{\text{bind}}$ (or $\frac{1}{K_D})$ is computed as:
\begin{equation}
    K_{\text{bind}} = \int_0^{r_c} 4\pi r^2 e^{-\frac{PMF(r)}{k_BT}} dr 
    \label{eqn:pmf-to-Kbind}
\end{equation}
where $r_c$ is the cut-off distance for the unbinding event. The standard binding free energy equivalent to Equation \ref{eqn:delta-G} is given by:
\begin{equation}
    \Delta G^{\circ} = k_BT \ln(K_{\text{bind}} C^0)
    \label{eqn:pmf-to-delta-G}
\end{equation}
For standard concentration, the value of $C^0$ is $(1/1660)$ {\AA}$^{-3}$. \cite{Souza2020ProteinligandModel,Gilson1997TheReview}}

Recent developments in milestoning theory opened up the possibility of calculating the committor distribution along the transition coordinate \cite{Elber2017CalculatingMilestoning}. The committor $C_i$ of a milestone $i$ is defined as the probability of a trajectory starting from milestone $i$ to eventually reach the product milestone before going to the reactant milestone. To calculate committor values the transition kernel $\mathbf{K}$ has to be modified with such a boundary condition that trajectories reaching the product milestone will "stick" there and will not return to previous milestones \cite{Elber2017CalculatingMilestoning,Ma2017FreeFlipping}. For a 3 milestone case this can be illustrated as: 

\begin{equation}
    \mathbf{K} = 
    \begin{bmatrix} 
    0 & K_{12} & 0\\
    K_{21} & 0 & K_{23}\\
    0 & K_{32} & 0
    \end{bmatrix} \rightarrow \mathbf{K_C} =
    \begin{bmatrix} 
    0 & K_{12} & 0\\
    K_{21} & 0 & K_{23}\\
    0 & 0 & 1
    \end{bmatrix}
\end{equation}
The committor vector $\mathbf{C}$ is then calculated as:
\begin{equation}
    \mathbf{C} = \lim_{n \to \infty} (\mathbf{K_C})^n \mathbf{e_p}
\end{equation}
where $\mathbf{e_p}$ is a unit vector whose all elements are zero except for the one corresponding to the final milestone. Numerically, multiple powers of $\mathbf{K_C}$ are computed until the committor converges. We considered the committors converged when the change in the norm of the $\mathbf{C}$ vector is less than $10^{-3}$. 

We implemented WEM using the WESTPA software \cite{Zwier2015} and the colvars module of the NAMD2 \cite{Phillips2005} molecular dynamics package. All trajectories are analyzed through the $w\_ipa$ module of the WESTPA code. 

\subsection{Error analysis}
There are multiple ways to perform error analysis for the WE and milestoning methods. We calculated the uncertainties in our results by generating an ensemble of transition matrices sampled from a Bayesian type conditional probability described in Ref. \citen{Majek2010,Vanden-Eijnden2008} and \citen{Votapka2015MultiscaleMilestoning}. We described this method in detail in our previous work \citen{Ray2020WeightedSimulations}. A brief outline is provided here.

The rate matrix $\mathbf{Q}$ is defined as 
\begin{equation}
\begin{split}
    &Q_{ij} = \frac{K_{ij}}{\overline{T}_i} \;\;\;\; (i \ne j) \\
    &Q_{ii} = - \sum_{i \ne j} Q_{ij}
\end{split}
    \label{eqn:qij_wem}
\end{equation}

Given a set of transition counts and lifetimes the probability of obtaining the matrix $\mathbf{Q}$ is given by:
\begin{equation}
    p(\mathbf{Q}|\lbrace N_{ij},\overline{T}_{i} \rbrace ) = \prod_i \prod_{j \neq i} Q_{ij}^{N_{ij}} \exp{(-Q_{ij} N_i \overline{T}_i)} P(\mathbf{Q})
    \label{eqn:likelihood}
\end{equation}
where $P(\mathbf{Q})$ is a uniform prior, $N_i$ is the total number of trajectories starting from milestone $i$, out of which $N_{ij}$ trajectories hit milestone $j$. The $\mathbf{Q}$ matrices are sampled from the distribution in Equation \ref{eqn:likelihood} using a non-reversible element exchange Monte-Carlo scheme \cite{Noe2008ProbabilityModels}. In each step one randomly chosen element and the diagonal element of the corresponding row of $\mathbf{Q}$ are updated as 
\begin{equation}
    \begin{split}
        &Q_{ij}' = Q_{ij} + \Delta \;\;\;\; (i \ne j)\\
        &Q_{ii}' = Q_{ii} - \Delta.
    \end{split}
    \label{eqn:q_move}
\end{equation}
where $\Delta$ is a random number sampled from an exponential distribution of range $[-Q_{ij},\infty)$ with mean zero. The new matrix $\mathbf{Q'}$ is accepted with a probability:
\begin{equation}
    p_{\text{accept}} = \frac{p(\mathbf{Q'}|\lbrace N_{ij},\overline{T}_{i} \rbrace )}{p(\mathbf{Q}|\lbrace N_{ij},\overline{T}_{i} \rbrace )} = \left( \frac{Q_{ij} + \Delta}{Q_{ij}}\right)^{N_{ij}} \exp{(-\Delta N_{i} \overline{T}_i )}
\end{equation}
As only one element is modified in one step, the sampled matrices are highly correlated \cite{Noe2008ProbabilityModels}. So we performed 10000 sampling steps out of which some changes are accepted and the others rejected. We only selected the matrices every 500 steps for our analysis. This provided us with 20 distinct $\mathbf{Q}$ matrices. From each $\mathbf{Q}$ the transition kernel $\mathbf{K}$ and lifetime vector $\mathbf{\overline{T}}$ are extracted as 
\begin{equation}
    \begin{split}
        &K_{ij} = \frac{Q_{ij}}{\sum_{l \in \lbrace i-1,i+1 \rbrace} Q_{il}  } \;\;\;\; (i \ne j) \\
        &\overline{T}_i = \frac{1}{\sum_{l \in \lbrace i-1,i+1 \rbrace} Q_{il}}
    \end{split}
    \label{eqn:QKT}
\end{equation}
Then all the calculations described in Section \ref{sec:WEMscheme} are performed on each set of $\mathbf{K}$ and $\mathbf{\overline{T}}$. The mean and 95\% confidence interval on these datasets have been reported throughout the manuscript unless mentioned otherwise. 

\section{Computational Methods}
\label{sec:methods}
\subsection{Systems studied:}
In this work we studied the association of sodium and chloride ion in solution as well as the binding of the 4-hydroxy-2-butanone (BUT) ligand to FK506 binding protein (FKBP). Here we give the simulation details for both systems.

\subsection{NaCl in water}
For our studies of association of sodium chloride in water, all molecular dynamics simulations were performed using NAMD 2.13 package \cite{Phillips2005}. A 40 \AA $\times$40 \AA $\times$40 \AA$ $ water box was prepared with TIP3P water \cite{Jorgensen1983}. A Na$^+$ ion was placed at the center of the box. A Cl$^-$ ion was placed along the x-axis at the following distances from the Na$^+$: 2.45 \AA, 2.7 \AA, 3.6 \AA, 4.6 \AA, 5.6 \AA, 7.0 \AA, 9.0 \AA, 11.0 \AA, and 13.0 \AA. These are the position of the milestones in this study. CHARMM36 force field \cite{MacKerell1998All-atomProteins} was used to model the sodium and chloride ions. WE bins were place in between milestones at 0.05 \AA$ $ interval between 2.45 \AA$ $ and 2.7 \AA, 0.2 \AA$ $ interval between 2.8 \AA$ $ and 3.6 \AA$ $, and 0.5 \AA$ $ interval afterwards except for a 0.4 \AA$ $ interval between 5.6 \AA, and 6.0 \AA. At distances above 6\AA, all bins were 0.5 \AA$ $ apart. Our choice of milestone positions was influenced by previous studies which suggest that the bound state minima of Na+/Cl- ion pair is at 2.7 \AA$ $ distance and the transition state of dissociation is at 3.6 \AA$ $ \cite{Timko2010DissociationSimulations}. Initially all the structures were minimized for 10000 steps using conjugate gradient algorithm, followed by an equilibration of 200 ps, with fixed ion positions. A time step of 1 fs was used. All simulations were performed in the NPT ensemble. The temperature was kept constant at 298 K using a Langevin thermostat with coupling constant of 5 ps$^{-1}$, and the pressure was maintained constant at 1 atm using a Langevin barostat \cite{Feller1995ConstantMethod}. Particle-mesh Ewald summation \cite{Essmann1995AMethod} was used for electrostatics, with a real space cutoff of 12 \AA$ $ for the non-bonded interactions. 

All WEM trajectories were started from the endpoints of the equilibration simulations. Five trajectories were propagated in each of the occupied bins. Multiple (500-2000) iterations were performed with $\delta t$ = 0.1 ps until results converged. Convergence plots for $K_{ij}$ values are shown in the supplementary information. Trajectories were split and merged if they crossed the bin boundaries after each iteration, preserving the total probability. The weights and propagation times of the trajectories reaching either of the nearest milestones were recorded and analyzed according to the WEM scheme (Sec. \ref{sec:WEMscheme}).

Because stopping a trajectory exactly at the milestone interface \cite{Bello-Rivas2015} is difficult in the current framework, we allow the trajectories to propagate until the end of the iterations. This incurs small additional computational costs but avoids the need to externally monitor the simulation at every MD time step. We do not split or merge the trajectories after crossing the milestones and we exclude the excess part of the trajectory from our analysis. This is not a problem as we are not doing exact milestoning which requires restarting trajectories from the same phase space point at which a previous trajectory has crossed a milestone. In short, we performed a single iteration of the exact milestoning but accelerating the convergence of transition kernel and lifetime matrix by applying WE.

The stationary flux, free energy profile and mean first passage time (MFPT) were computed using equation \ref{eqn:statflux}, \ref{eqn:free-energy}, and \ref{eqn:MFPT} respectively. MFPT of the reverse process was also calculated from which information about $k_{\text{on}}$ can be obtained. To validate our results a 300 ns long equilibrium MD simulation was performed starting from the bound state of NaCl with identical simulation parameters. The distance between Na$^+$ and Cl$^-$ is stored every 500 fs and the free energy is computed from the distribution. Mean first passage times and rates ($k_{\text{on}}$ and $k_{\text{off}}$) were computed from 10 dissociation and association events observed during the simulation. The sampling is not exhaustive but provides an order of magnitude estimate of the binding and unbinding kinetics.

\subsection{FKBP-BUT complex}
To study a protein-ligand system, we focused on binding and unbinding of the BUT ligand to the FKBP protein. We obtained the initial structure from the crystal structure in the RCSB PDB database (PDB ID: 1D7J) \cite{Burkhard2000X-rayEnergies}. The protein is modelled by CHARMM36m force field with CMAP correction \cite{Huang2017}. The parameters for ligand were generated using CGenFF \cite{Vanommeslaeghe2012AutomationTyping,Vanommeslaeghe2012AutomationCharges}. The protein was solvated in a water box of size 89 \AA $\times$ 68 \AA $\times$ 76 \AA$ $ to provide adequate space for the ligand to release. The additional space was also needed to observe the convergence of the free energy profile of the ligand as a function of distance from the protein. One Cl$^-$ ion was added to the system to neutralize the charge (following Ref. \citen{Dickson2016LigandMechanisms}) as necessary for the particle mesh Ewald (PME) method. All input files were generated using the CHARMM-GUI web server \cite{Jo2008} and Psfgen package of VMD \cite{Humphrey1996}. All simulations were performed using NAMD 2.13 \cite{Phillips2005}.

The ligand bound FKBP was first minimized by 50000 steps using conjugate gradient algorithm followed by a short equilibration with gradually decreasing harmonic restraint on the protein and ligand heavy atoms over 600 ps. The system is further equilibrated for 1 ns in NPT ensemble with no restraint on heavy atoms. The time step for equilibration was kept at 1 fs, bonds to hydrogen atoms were constrained using SHAKE \cite{Ryckaert1977NumericalN-alkanes} algorithm. From the final structure, a production run, with 2 fs time step, is performed for 20 ns to generate a ligand unbinding trajectory. From this trajectory the anchors (seed structures) for individual milestones were obtained. For all simulations, a Langevin thermostat with coupling constant 1 ps$^{-1}$ is used to maintain the temperature at 298 K. A Langevin barostat kept the pressure constant at 1 atm. The simulation parameters were identical for all further simulation.

The reaction coordinate for ligand dissociation was chosen to be the distance ($r$) between the center of mass of the protein and the ligand, following previous work \cite{Huang2011TheUnbinding}. The bound state from crystal structure is characterized by $r \approx$ 6 \AA. Three trials of independent WEM simulations were performed. Milestones were placed at $r$ = 5 \AA$ $ and at 2 \AA$ $ interval between $r$ = 6 \AA$ $ and 28 \AA. Starting structures for each protein-ligand distance (milestone) was sampled from the ligand release trajectory. Each structure was equilibrated for 100 ps with strong harmonic restraint along the reaction coordinate ($r$) with a force constant gradually increasing up to 80 kcal mol$^{-1}$\AA$^{-2}$ over the first 50 ps and kept constant for the rest 50 ps. The final frame was used to propagate WE simulation. Due to the stochastic nature of the thermostat, repetition of this exact same procedure 3 times generates 3 different sets of WEM input structures per milestone, which we used for three trials. 

We tried different milestone spacings (results not shown in this paper) but 2 \AA$ $ spacing is found to be optimum. \textcolor{black}{The trajectories need to lose the memory of their starting point before reaching a different milestone. The loss of memory can be quantified by calculating the velocity autocorrelation function of the reaction coordinate. All the trajectories starting from any given milestone should spend longer than the de-correlation time before reaching another milestone. In supplementary material, we show that a 2 {\AA} spacing in milestones satisfies this criteria.} If a shorter interval is used, the trajectories fail to lose memory of the starting structure while a larger spacing provides a very sparse free energy profile which is hard to interpolate. WE bins were placed at 0.5 \AA$ $ interval, and each iteration of WE calculation has $\delta t = 2$ ps. The bin width and $\delta t$ pair satisfies the condition that a fraction of the trajectories cross the bin boundary in about one iteration \cite{Bogetti2019AV1.0}. 
The number of trajectories per WE bin was kept constant at 5. 

To compare the residence time and $k_{\text{off}}$ results, 20 independent conventional MD trajectories,  with different starting velocities, were propagated starting from the equilibrated bound FKBP-BUT complex. Each trajectory was propagated for 20 ns. Out of them 17 trajectories showed ligand release events. The mean and 95\% confidence interval was calculated for the residence time from those trajectories and $k_{\text{off}}$ was computed using equation \ref{eqn:koff-kon}. \textcolor{black}{All the previous studies with this system used either a different force field or  different reaction coordinates and distance criteria for binding/unbinding. To make a fair comparison with the binding free energy data from WEM, a well-tempered metadynamics simulation was performed starting from the equilibrated bound state structure and with simulation parameters identical to those used in the rest of the paper. The free energy profile was obtained as a function of protein-ligand distance and the PMF was computed by correcting for the radial Jacobian factor. From the PMF the binding free energy was calculated using Equations \ref{eqn:pmf-to-Kbind} and \ref{eqn:pmf-to-delta-G}. Further details about the metadynamics simulation are provided in the Supplementary Material.}

\subsection{Harmonic Restraint Release}
We obtained very high variability of the free energy and $k_{\text{off}}$ values for FKBP-BUT complex using our standard WEM scheme \cite{Ray2020WeightedSimulations} which uses only one starting structure per milestone for each independent WEM run. We realized the necessity to have multiple starting structures for obtaining appropriate statistics although we can get multiple hitting points from a single starting structure. One way to achieve that is to perform multiple rounds of WEM simulation. The combined statistics from independent runs for each individual milestone could then be used to construct the transition kernel ($\mathbf{K}$) and lifetimes ($\mathbf{\overline{T}}$). This process was performed for the three independent WEM trials and all the observables were computed using the combined (or averaged) $\mathbf{K}$ and $\mathbf{\overline{T}}$ using milestoning protocol described in Sec. \ref{sec:WEMscheme}. The results of this exercise are referred to as "WEM average". 

\begin{figure}
    \centering
    \includegraphics[scale=0.4]{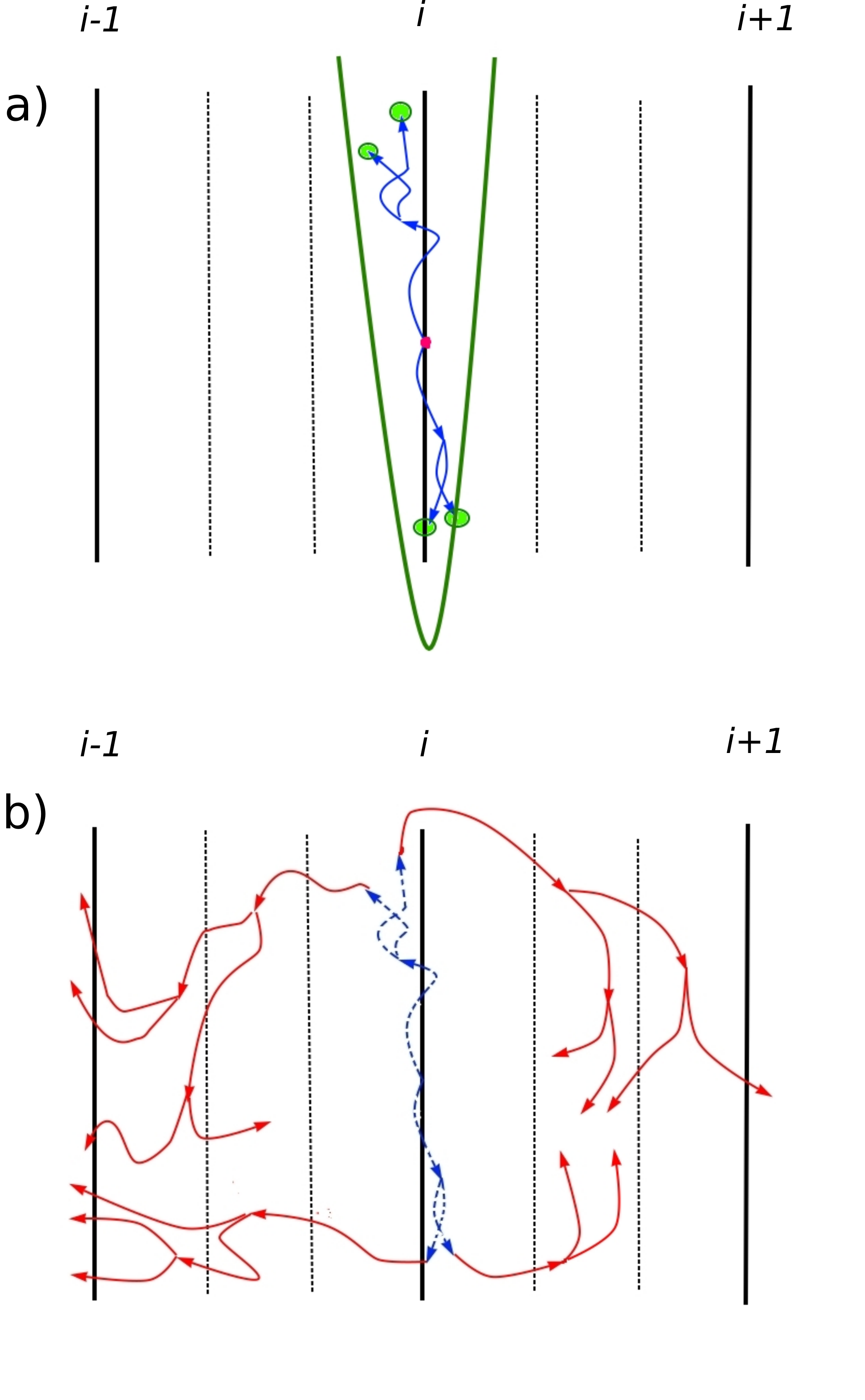}
    \caption{\textcolor{black}{Schematic diagram of the weighted ensemble milestoning with restraint release (WEM-RR). (a) MD trajectories initiated from milestone ($i$) anchor (pink filled circle).  The harmonic restraint confines the dynamics to milestone $i$, but fluctuations "inside" the interface spawn new trajectories, leading to multiple starting structures (green filled circles). Only two WE iterations of restrained dynamics are depicted. (b) When restraint is released, trajectories propagate via regular WE (red solid line) until they reach either milestone $i-1$ or $i+1$. The restrained part of the dynamics (blue dashed line) is discarded from the analysis. Effectively, we obtain a set of WEM transitions starting from four different starting points on milestone $i$. In the schematic example, trajectories have probability of 7/16 when reaching milestone $i-1$ and 1/16 for milestone $i+1$. The sum of the probabilities converges to 1 after sufficient sampling.}}
    \label{fig:schematic}
\end{figure}

But this did not improve our results significantly, as described in section \ref{sec:results}. Increasing the number of starting structures to a large extent is not practical because in the weighted ensemble method multiple daughter trajectories are generated through splitting of a small number of initial parent trajectories. Another possibility is to discard the first few iterations of the WEM trajectories for each milestone. In this way we get a set of trajectories with different starting points. But most of such trajectories have their starting points in a bin which is far from the milestone interface. So most of the simulation effort in the initial iterations would be wasted in generating irrelevant starting structures. To address this, we used a harmonic restraint around the milestone interface and performed a few (10 in this case) WE iterations starting from one starting structure. As we allow 5 trajectories per bin, we eventually get total 10 trajectories, 5 slightly left of the milestone and 5 slightly to the right. This is a result of the fact that a WE bin boundary coincides with the milestone interface. We chose the harmonic restraint to be 50 kcal mol$^{-1}$ \AA$^{-2}$ for all milestones except for the one at $r=$ 5 \AA, which had 100 kcal mol$^{-1}$ \AA$^{-2}$. 
The advantage of this technique is that, even in the presence of a very steep energy landscape, we get an equal number of starting points on both sides of the milestone. This almost never happened in the standard WEM protocol and the one starting point is generally on the side of the lower free energy. Having an equal number of points on both sides does not cause any errors in probability calculation as these structures will have different weights to begin with. This modification, which we call weighted ensemble milestoning with restraint release (WEM-RR), has been applied to study the FKBP-BUT complex. 
Both procedures were initiated from the same structures on each milestone in order to measure the benefit gained from the applied restraint-release step.
The averaging of the WEM-RR results were also performed for three independent trials, following the same protocol used for the regular WEM method, and the results were reported as "WEM-RR average".

\section{Results}
\label{sec:results}
\subsection{NaCl in water}

\begin{figure}
\centering
\includegraphics[width=0.5\textwidth]{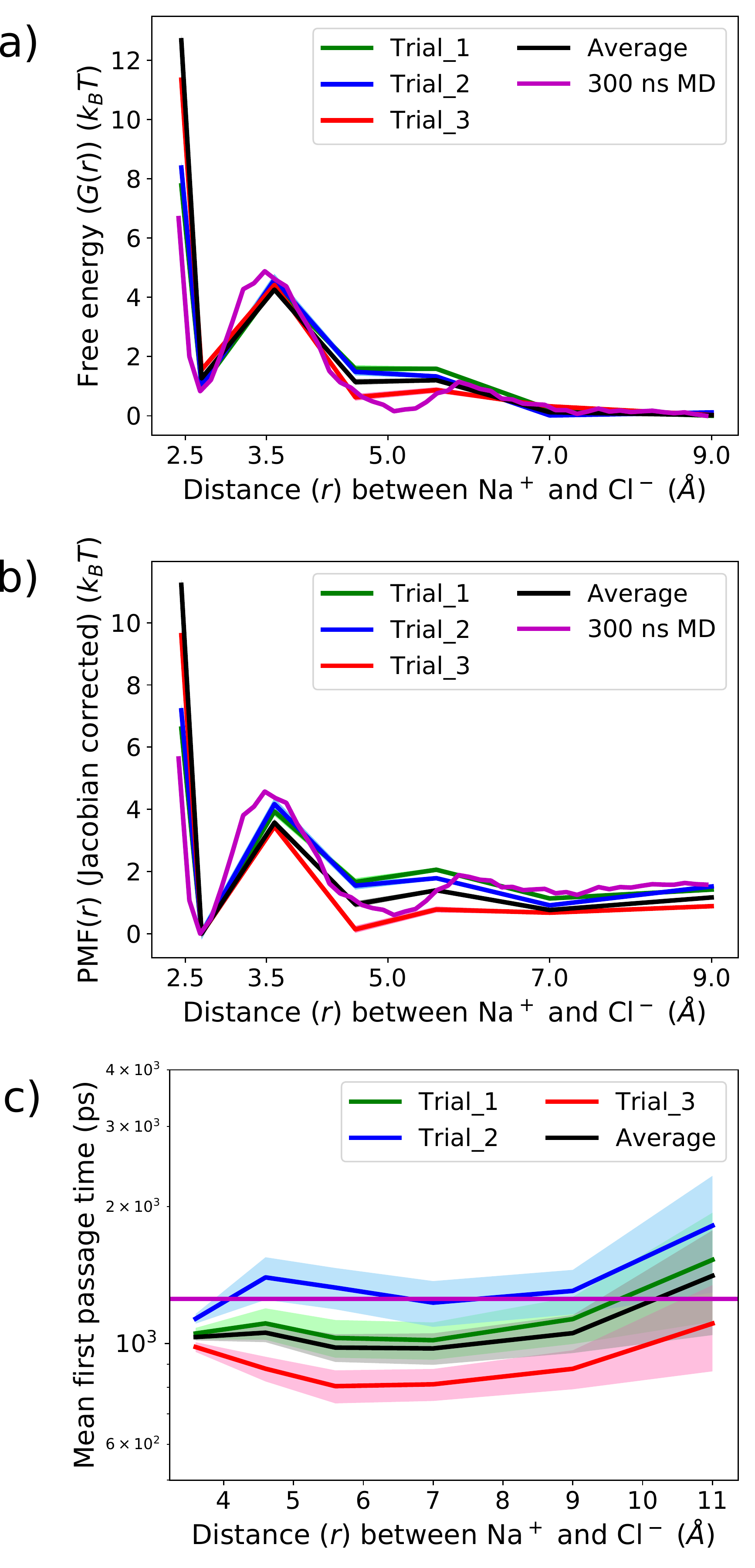}
\caption{(a) Free energy profile along the distance between Na$^+$ and Cl$^-$ computed using WEM simulation and long equilibrium MD simulation. (b) A PMF for the same as top with appropriate Jacobian correction. (c): Mean first passage time to each of the milestones starting from the bound state ($r = 2.7$\AA). The horizontal line is the residence time computed from long equilibrium MD simulation.}
\label{fig:NaCl}
\end{figure}

Figure \ref{fig:NaCl} shows the free energy profiles as a function of Na$^+$ to Cl$^-$ distance ($r$) computed from both conventional MD and WEM simulation. For conventional MD the value of $r$ was sampled from a 300 ns MD simulation and the free energy profile was computed from the histogram of the data. There is a deep minimum at $r = 2.7$ \AA$ $ indicating the bound state, which is followed by a sharp free energy barrier at $r=3.6$ \AA. At a slightly larger separation $r = 5.2$ \AA, there is a shallow minimum indicating the formation of a second hydration shell between the ion pair \cite{Wolf2018TargetedCorrection,Mullen2014TransmissionDissociation}. The free energy profile is flat after a separation of 7 \AA$ $ and is devoid of any interesting features. So we define a state to be unbound when $r > 7$ \AA$ $ and bound when $r < 3.6$ \AA$ $. 

The free energy profile obtained from all three sets of weighted ensemble milestoning (WEM) simulation shows excellent agreement with the results from conventional MD. The free energy minima and the barrier heights could be estimated with $<$1 $k_BT$ accuracy (Figure \ref{fig:NaCl}). The Jacobian corrected PMF for the system has also been calculated using Equation \ref{eqn:jacobian} as shown in figure \ref{fig:NaCl}. Small structural features like the formation of second hydration shell is not very apparent from the WEM free energy profiles because in milestoning based methods the free energy can only be calculated at the milestone interfaces, leading to values of the PMF being defined only at intervals which are well separated, making it hard to distinguish the finer structure. To obtain more detailed features of the PMF we would need to place milestones at smaller intervals. However, the interval must be large enough such that the process remains memoryless. Because of this, there is an effective limit to the level of detail the PMF can represent. If finer details of the ruggedness in between milestones are needed, non-milestoning methods (such as umbrella sampling or other biased methods) can be used locally.
 

We computed the mean first passage times (MFPT) for transitions from the bound milestone (minimum at $r = 2.7$ \AA) to all other milestones (Fig. \ref{fig:NaCl}). The results of three trial WEM runs agree with each other. The unbinding time computed from the continuous MD trajectory is in accord with 
the MFPT from milestone at $r = 2.7$ \AA$ $ to $r = 7.0$ \AA$ $ calculated using WEM method (Table \ref{tab:NaCl}). Since we only observed 10 transition events between the bound and unbound states in the 300 ns conventional MD simulation, the residence time ($\langle \tau \rangle_{\text{off}}$) has high uncertainty (Table \ref{tab:NaCl}).

Using the above comparison, the binding time ($\langle \tau \rangle_{\text{on}}$) from the WEM simulations were computed as the mean first passage time for the transition from milestone at $r = 7.0$ \AA$ $ to the bound interface ($r = 2.7$ \AA). The average binding and unbinding times were found to be $\sim$37 ns and $\sim$1 ns (Table \ref{tab:NaCl}). Compared to the conventional simulation, the unbinding time showed good agreement, but the binding times differed by an order of magnitude. We believe this is primarily because of the poor statistics of binding events captured in the conventional simulation.

\begin{table*}
\caption{\label{tab:NaCl} Kinetic and thermodynamic properties for the dissociation of Na+/Cl- ion pair. All error bars are 95\% confidence intervals. The averaging for WEM results was performed at individual milestone level and the observables were calculated using the averaged $\mathbf{K}$ and $\mathbf{\overline{T}}$ in the milestoning framework. }
\begin{ruledtabular}
\begin{tabular}{ccccccc}
Simulation &$\langle \tau \rangle_{\text{off}}$  &$\langle \tau \rangle_{\text{on}}$ 
&$k_{\text{off}}$  &$k_{\text{on}}$ \footnote{$v = 58500$ \AA$^3$} &$K_D$  &$\Delta G^{\circ} $\footnote{Estimated using Eqn. \ref{eqn:delta-G}} \\
 &(ns) &(ns) &$(\times 10^6 s^{-1})$ &$(\times 10^9 M^{-1}s^{-1})$ &(M) &$(k_BT)$ 
\\
\hline
\hline
WEM trial 1 &1.02$\pm$0.09 &30.5$\pm$5.00 &910$\pm$87 &1.2$\pm$0.19 &0.8$\pm$0.14 &0.2$\pm$0.2 \\
WEM trial 2 &1.23$\pm$0.14 &57.9$\pm$8.78 &810$\pm$93 &0.61$\pm$0.09 &1.3$\pm$0.25 &-0.3$\pm$0.2 \\
WEM trial 3 &0.81$\pm$0.07 &45.0$\pm$3.55 &1200$\pm$110 &0.78$\pm$0.06 &1.5$\pm$0.18 &-0.4$\pm$0.1\\
WEM average &0.97$\pm$0.08 &37.3$\pm$5.04 &1030$\pm$85 &0.90$\pm$0.13 &1.1$\pm$0.19 &-0.1$\pm$0.2\\
Long MD &1.25$\pm$0.81 &285.8$\pm$193.6 &800$\pm$520 &0.12$\pm$0.08 &7.0$\pm$6.2 &-1.9$\pm$0.9\\

\end{tabular}
\end{ruledtabular}

\end{table*}

We used Eq. (\ref{eqn:koff-kon}) to estimate the unbinding and binding rate constants $k_{\text{off}}$ and $k_{\text{on}}$, respectively. The volume of the equilibrated water box fluctuated between 58000 \AA$^3$ and 59000 \AA$^3$ in the constant pressure simulation. We used the average volume $v = 58500$ \AA$^3$ in Equation \ref{eqn:koff-kon}.  The values of $k_{\text{off}}$ and $k_{\text{on}}$ estimated from WEM simulation are $10^9$ s$^{-1}$ and $10^9$ M$^{-1}$s$^{-1}$ respectively. The mean first passage times and rate constants are within one order of magnitude of those obtained from  NaCl dissociation and association studies using AMBER force field \cite{Votapka2015MultiscaleMilestoning,Wolf2018TargetedCorrection,Wolf2020MultisecondSimulations}. The WEM simulations produced a dissociation constant $K_D$ close to 1 M and a binding free energy $\sim$ 0 (using Equation \ref{eqn:KD}). These numbers are within 1 $k_BT$ of the $\Delta G$ values calculated from separate classical CHARMM and ab initio simulations reported by Timko et al \cite{Timko2010DissociationSimulations}.

In essence, the application of the WEM method to the dissociation of the Na+/Cl- ion pair produced a free energy profile and residence time consistent with conventional MD simulation. Additionally, the binding time, rate constants, and dissociation constant could also be calculated.

\subsection{FKBP-BUT complex}
Next, we apply WEM to a protein/ligand unbinding process with a relatively complex energy landscape. Although not as simple as the case of NaCl, we show that WEM-RR is still able to produce results comparable to expensive, extended MD simulations at a fraction of the computational cost.

\begin{figure*}
\centering
\includegraphics[width=\textwidth]{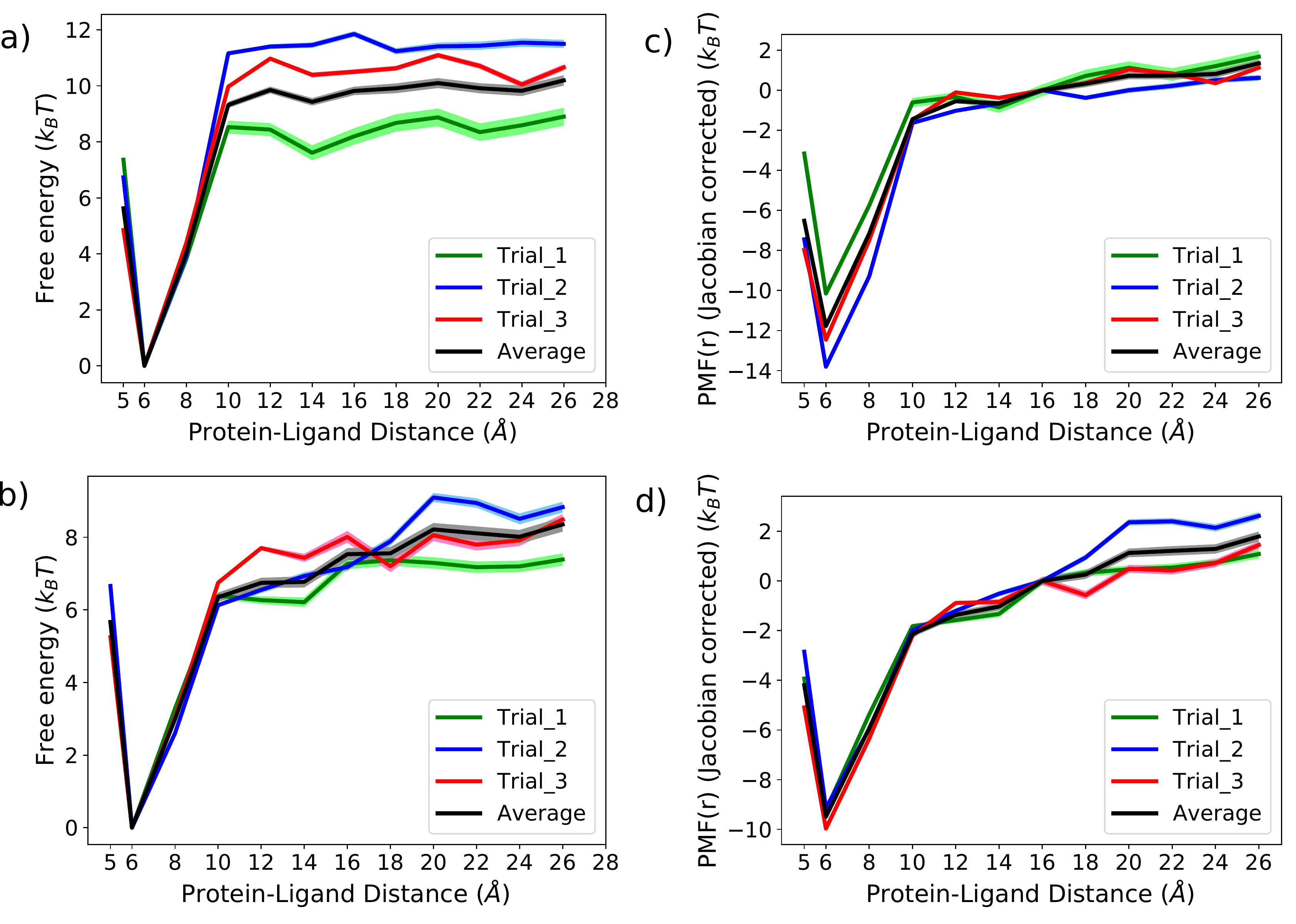}

\caption{Free energy profile along the center of mass distance between FKBP protein and BUT ligand: (a) using regular WEM calculation and (b) using harmonic restraint release (WEM-RR) to generate multiple initial conformations per milestone. The PMFs obtained from the free energy profile for (c) WER and (d) WEM-RR simulation are in the right column. The PMF scale was chosen to be zero at $r=16$ {\AA}.
}

\label{fig:WEM-freeE}
\end{figure*}

The binding thermodynamics and kinetics of the ligand BUT to the enzyme FKBP were studied using WEM sampling along the center of mass separation. The free energy profile of BUT ligand dissociation from the FKBP protein is illustrated in Fig. \ref{fig:WEM-freeE}. All trials showed a clear minimum corresponding to the bound state at about 6 \AA$ $. This is consistent with the protein-ligand distance in the crystal structure. The free energy converges to a constant value after 14 \AA, with fluctuations $< 1 k_BT$, indicating that the ligand is unbound, with little influence of the protein on the ligand. From this observation, we used the milestone at 16 {\AA} as the point where BUT is considered unbound for the purposes of calculating kinetics, specifically $k_{\rm off}$. Similarly for calculating residence time we report the mean first passage time of transition from milestone at $r = 6$ \AA$ $ to milestone at $r = 16$\AA$ $. \textcolor{black}{The value of $r_c$ in Equation \ref{eqn:pmf-to-Kbind} for integrating the PMF's were also chosen to be $r = 16$\AA$ $.}

\begin{figure}[h!t]
\centering
\includegraphics[scale=0.5]{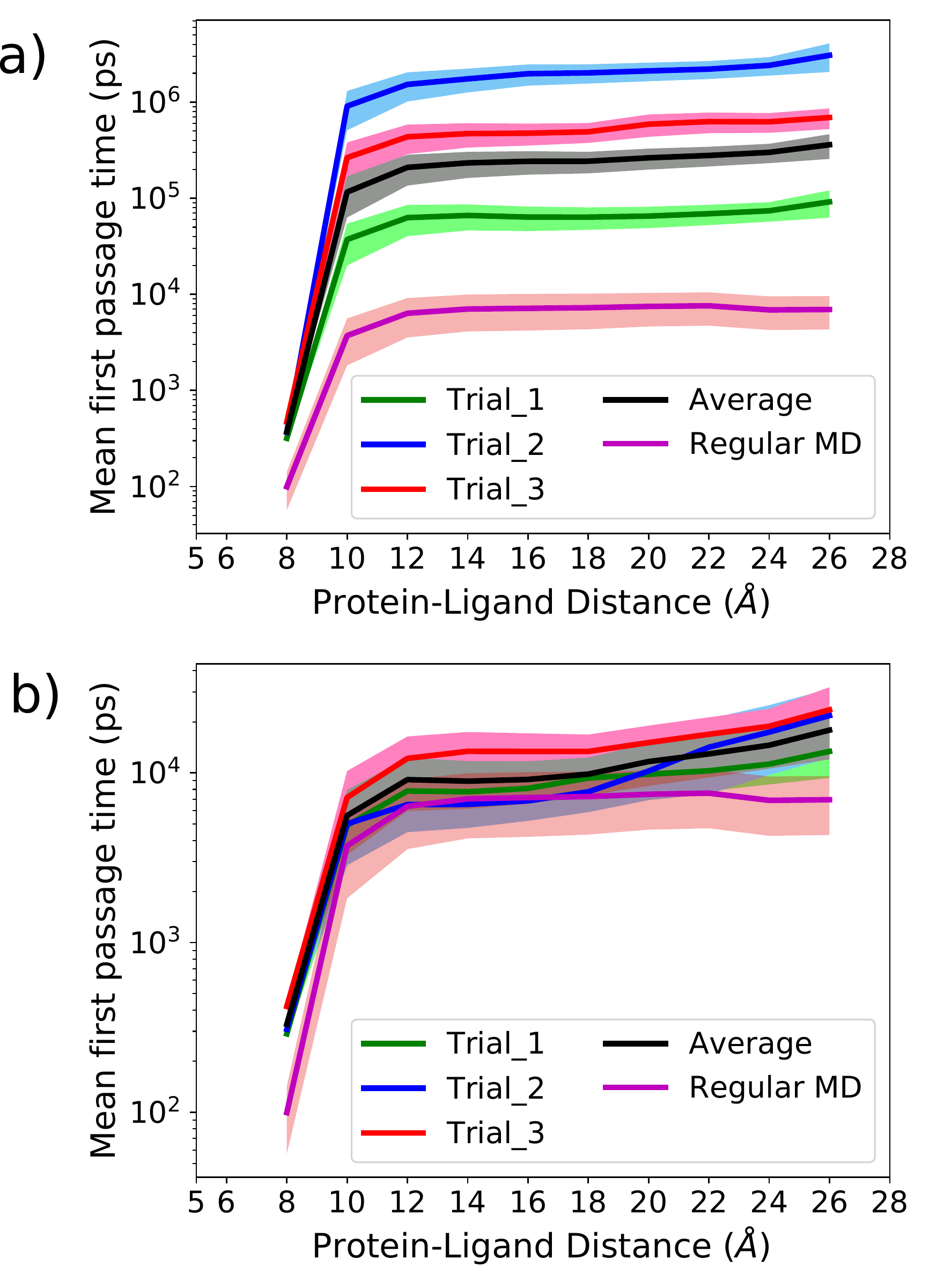}
\caption{Mean first passage time to different milestones starting from the bound state r=6 \AA: (a) using normal WEM calculation and (b) using harmonic restraint release to generate multiple initial condition per milestone. The result obtained from 20 regular MD trajectories has also been depicted.}
\label{fig:MFPT}
\end{figure}

The free energy profiles calculated from three independent WEM trials showed a large variance of binding free energy between 8-12 $k_BT$. Similarly, for the ligand residence time in Fig. \ref{fig:MFPT} we observed results spanning over two orders of magnitude. To investigate these issues with the aim to increase the accuracy of the WEM method, we focused on ways to improving the initial conditions of the method. Since the WEM approach is based on the approximate form of milestoning \cite{West2007}, compared to the much more expensive exact approach \cite{Bello-Rivas2015}, we found that our calculations were sensitive to the starting coordinates defined on each milestone. To alleviate the sampling issues, we show that inclusion of a "restraint-release" cycle prior to WEM application allows for a more thorough sampling of initial conditions (see Section \ref{sec:methods} for details). 

We generated multiple starting points per milestone using harmonic restraint release. These starting points were used to propagate unbiased WEM trajectories for further analysis. This approach significantly reduces the variance of the data both for free energy profiles (see Fig. \ref{fig:WEM-freeE}) and for the residence time (Fig. \ref{fig:MFPT}). The 10 starting structures sampled from the harmonically restrained first few WE iterations for one of the milestones are shown in Supplementary Material. Multiple distinct orientations of the ligand were sampled within this 10 starting structures, which enhanced the accuracy of the milestoning calculation.

The residence time and unbinding rate constants were calculated using WEM and WEM-RR simulations and compared against the average value of 20 conventional MD trajectories. Mean first passage time to all milestones have been calculated and depicted in figure \ref{fig:MFPT}. The results of WEM-RR shows significantly better agreement with the conventional MD results compared to the regular WEM which overestimated the timescales by a few orders of magnitude. \textcolor{black}{The binding free energy was also computed from the PMF's obtained from WEM and WEM-RR simulation (Fig. \ref{fig:WEM-freeE}) using Equation \ref{eqn:pmf-to-Kbind} and \ref{eqn:pmf-to-delta-G}. The results were compared against the $\Delta G^0$ from the PMF obtained from well tempered metadynamics simulation (see supplementary material).   }

Comparing the results with previous simulation studies should be done carefully because different force field, reaction coordinate and distance criteria for unbinding have been used. The residence time and binding free energy of our work (WEM-RR) agrees better with the results of Huang and Caflisch, and Lotz and Dickson; both groups used  a CHARMM force field in their simulation. Additionally, the ligand residence time from our work is within the same order of magnitude with that of the AMBER ff99SB*-ILDN + GAFF simulations \cite{Pramanik2019CanDissociation,Pan2017QuantitativeSimulations} (Table \ref{tab:MFPT}). Our reaction coordinate (RC) is similar to the binding-pocket to ligand distance used by Huang and Caflisch, but different from the RCs used in other studies. For example, Dickson and Lotz used a ligand RMSD based RC, Pan et al. used the distance between between Trp59 and the ligand, while Pramanik et al. constructed a novel RC based on a linear combination of multiple inter-atomic distances involving the protein and the ligand. Despite the different definitions for RC, our kinetic results are consistent with the existing literature.  

Other important properties of ligand-receptor interaction, such as $k_{\text{off}}$, $k_{\text{on}}$, $K_D$,  and $\Delta G$, were calculated and the results are reported in Table \ref{tab:MFPT}. The value of $\langle \tau \rangle_{\text{on}}$ was calculated as the MFPT of the transition from milestone $r = 16$\AA, to milestone $r = 6$\AA. The $k_{\text{on}}$ was then obtained using equation \ref{eqn:koff-kon}. The binding affinity $K_D$ is obtained as a ratio of $k_{\text{off}}$ and $k_{\text{on}}$. The $\langle \tau \rangle_{\text{off}}$ and $k_{\text{off}}$ values computed from WEM-RR agree very well with the conventional MD results. It is difficult to compare the numbers for $\langle \tau \rangle_{\text{on}}$ and $k_{\text{on}}$ as they are relatively hard to obtain from direct MD simulation because of the large computational cost as demonstrated for the Na+/Cl- ion pair. 
As our residence times agree with that of conventional MD simulation as well as literature results and our $K_D$ ($k_{\text{off}}/k_{\text{on}}$) shows order-of-magnitude similarity with the experimental $K_D$, we can say with some degree of confidence that our binding times and rate constants are reasonably accurate. We observed a discrepancy by two orders of magnitude in our $K_D$ value with those obtained from equilibrium simulations by Pan et. al.\cite{Pan2017QuantitativeSimulations}. We address this issue and propose modifications to our current analyses in the next two paragraphs. In short, our results show that calculated kinetic parameters are sensitive to how the diffusive regime is modeled.

It can be argued that while the binding rate constant is often diffusion limited, the effect of diffusion is ignored when $k_{\text{on}}$ is computed from the transition timescale from one milestone to another. To the best of our knowledge there are no experimental data on $k_{\text{on}}$ for the FKBP-BUT system to compare to. The only available literature value is the result obtained by Pan et al.\, who use very different criteria for binding or unbinding, and at the same time exhibiting a very large variance. As a consistency check of the numbers, we calculated the MFPT to all other milestones from $r = 6$\AA$ $ and the MFPT to $r = 6$\AA$ $ from all other milestones. Using this information we calculated $k_{\text{on}}$ and $k_{\text{off}}$ for all milestones (considering every milestone with $r>6${\AA} one at a time as the cutoff for unbinding). The value of binding affinity $K_D$, computed as the ratio $k_{\text{off}}/k_{\text{on}}$, converges rapidly with increasing $r$ and remains invariant at $r>12$\AA$ $ (see Supplementary material). This indicates that the distance criterion for unbinding in our calculation is a matter of choice beyond 12 \AA, as the ratio of $k_{\text{on}}$ and $k_{\text{off}}$ will remain unchanged. All results we reported in Table \ref{tab:MFPT} uses the $r = 16$\AA$ $ as the distance criterion for unbinding.

\begin{table*}
\caption{\label{tab:MFPT} Comparison of different kinetic and thermodynamic properties for FKBP-BUT complex from the current study and literature. All error bars are 95\% confidence intervals unless mentioned otherwise. The averaging for WEM and WEM-RR results was performed for each individual milestone and the observables were calculated using the averaged $\mathbf{K}$ and $\mathbf{\overline{T}}$ in the milestoning framework (as discussed in Sec. \ref{sec:methods})}
\begin{ruledtabular}
\begin{tabular}{ccccccccccc}
Simulation &$\langle \tau \rangle_{\text{off}}$  &$\langle \tau \rangle_{\text{on}}$ 
&$k_{\text{off}}$  &$k_{\text{on}}$ \footnote{$v = 374500$ \AA$^3$} &$k_{\text{on}}^{\text{diff}}$ &$K_D$ &$K_D^{\text{diff}}$ &$\Delta G^{\circ} $\footnote{Estimated using Egn. \ref{eqn:delta-G}} &$\Delta G^{\circ,\text{diff}}$ &$\Delta G^{\circ}$\footnote{Obtained \textcolor{black}{by integrating the PMF using Equation \ref{eqn:pmf-to-Kbind} and \ref{eqn:pmf-to-delta-G}}} \\
 &(ns) &(ns) &$(\times 10^6) $ &$(\times 10^9)$ &$(\times 10^9)$ &($\mu$M) &(mM)  &$(k_BT)$ &$(k_BT)$ &$(k_BT)$ \\
 & & &($s^{-1}$) &($M^{-1}s^{-1}$) &($M^{-1}s^{-1}$) & & & & & 
\\
\hline
\hline
WEM trial 1 &63.7$\pm$18.2 &0.55$\pm$0.10 &16$\pm$4.5 &410$\pm$75 &- &40$\pm$13 &- &10.1$\pm$0.3 &- &9.3\\
WEM trial 2 &1976$\pm$491 &0.74$\pm$0.08 &0.5$\pm$0.13 &300$\pm$33 &- &1.7$\pm$0.5 &- &13.3$\pm$0.3 &- &12.9\\
WEM trial 3 &475$\pm$122 &0.81$\pm$0.07 &2.1$\pm$0.54 &280$\pm$24 &- &8$\pm$2 &- &11.7$\pm$0.2 &- &11.6\\
WEM average &242$\pm$66.4 &0.64$\pm$0.10 &4$\pm$1.1 &350$\pm$55 &- &11$\pm$3.6 &- &11.4$\pm$0.3 &- &10.9\\
WEM-RR trial 1 &8.11$\pm$2.38 &0.63$\pm$0.08 &120$\pm$36 &410$\pm$52 &1.1 &290$\pm$95 &109$\pm$33 &8.1$\pm$0.3 &2.2$\pm$0.3 &8.4\\
WEM-RR trial 2 &6.84$\pm$1.63 &0.45$\pm$0.02 &150$\pm$35 &500$\pm$22 &0.48 &300$\pm$71  &312$\pm$73 &8.1$\pm$0.2 &1.2$\pm$0.2 &8.3\\
WEM-RR trial 3 &13.4$\pm$3.73 &0.53$\pm$0.05 &70$\pm$21 &480$\pm$45 &0.56 &150$\pm$46 &125$\pm$37 &8.8$\pm$0.3 &2.1$\pm$0.3 &9.1\\
WEM-RR average &9.25$\pm$2.56 &0.53$\pm$0.04 &110$\pm$31 &420$\pm$35 &0.71\footnote{arithmetic average of three trials} &260$\pm$77 &170$\pm$26 \footnote{Calculated using the arithmetic average value of $\Delta G^{\circ,\text{diff}}$} &8.3$\pm$0.3 &1.8$\pm$0.2 &8.7\\
conventional MD &7.15$\pm$2.95 &- &140$\pm$58 &- &- &- &- &- &- &2.3\footnote{From well tempered metadynamics simulation}\\
\hline
Experiment \cite{Burkhard2000X-rayEnergies} &- &- &- &- &- &500 &- &7.6 &- &-\\
Huang \& Caflisch \cite{Huang2011TheUnbinding} &8$\pm$2 &- &- &- &- &- &- &7.33\footnote{calculated using linear interaction energy (LIE) model \cite{Hansson1998LigandMethods}} &- &-\\
Dickson \& Lotz \cite{Dickson2016LigandMechanisms} &4$\pm$1, 7$\pm$2 &- &- &- &- &- &- &- &- &-\\

Pramanik et. al \cite{Pramanik2019CanDissociation}. &21.3$\pm$0.2, 27.3$\pm$0.1 &- &- &- &- &- &- &- &- &2.0\\
Pan et. al. \cite{Pan2017QuantitativeSimulations} &- &- &45$\pm$4 &- &1.23$\pm$8 &- &38$\pm$4 &- &3.35$\pm$10.0 &-\\

\end{tabular}
\end{ruledtabular}

\end{table*}

The effect of diffusion in the $k_{\text{on}}$ values obtained from our work is worth consideration. We used an analytical model \cite{McCammon1986DiffusionalAssociation} to account for the effect of diffusion in the $k_{\text{on}}$ values. The method is described in detail in Appendix \ref{sec:diffusion}. The diffusion dependent arrival rate $k_{\text{on}}^{\text{diff}}$ on milestone $r = 16$\AA$ $ is two orders of magnitude smaller than the $k_{\text{on}}$ values computed for transition from $r=16$\AA$ $ to $r=6$\AA. It indicates that the arrival by diffusion is the rate determining step in ligand binding to FKBP. So we directly reported the $k_{\text{on}}^{\text{diff}}$ in table \ref{tab:MFPT} and did not try to combine these values with previously calculated $k_{\text{on}}$ numbers. The $k_{\text{on}}^{\text{diff}}$ values such obtained agree very well with the results obtained by Pan et al. \cite{Pan2017QuantitativeSimulations} who, in their simulation, allowed the ligand to diffuse around in the solvent after a dissociation event and included the time period of ligand diffusion into the $k_{\text{on}}$ calculation. We also calculated diffusion corrected binding affinity, $K_D^{\text{diff}}$, as the ratio $k_{\text{off}}/k_{\text{on}}^{\text{diff}}$ (Table \ref{tab:MFPT}). They are within an order of magnitude agreement with the $K_D$ values obtained long equilibrium simulation \cite{Pan2017QuantitativeSimulations}, although this number is far from the experimental value. The average binding affinity from diffusion corrected results ($\Delta G^{\circ,\text{diff}}$) is $1.8\pm0.2$ $k_BT$ (Table \ref{tab:MFPT}), which is within 2 $k_BT$ of the results of Pan et al. \cite{Pan2017QuantitativeSimulations} and within 1 $k_BT$ of Pramanik et al. \cite{Pramanik2019CanDissociation} \textcolor{black}{and the well tempered metadynamics results from our current work}.

If the effect of diffusion dependent arrival time is not included in the calculation, the $k_{\text{on}}$ we obtain, leads to a $K_D$ which is consistent with the experimental value, the binding free energy computed by Huang et al., and with the long distance limit of the WEM-RR free energy profile (this work, see Figure \ref{fig:WEM-freeE}). However, if the effect of diffusion is included (our calculations suggest that the binding process is diffusion dominant), our calculated $k^{\text{diff}}_{\text{on}}$ leads to the $K_D^{\text{diff}}$ results in Table \ref{tab:MFPT}. The resulting $\Delta G^{\circ, \text{diff}}$ using Eq. (\ref{eqn:delta-G}) is in quantitative agreement with our own metadynamics result and the results of Pan et al. and Pramanik et al. This $\Delta G$ of binding is however different by 5-6 $k_BT$ from the experimental value of binding affinity and the long distance limit of the free energy profile obtained from WEM-RR simulation (Fig. \ref{fig:WEM-freeE}). 
The values of $k_{\text{off}}$ and $\langle \tau \rangle_{\text{off}}$ are similar to the previous MD simulation studies with different force fields. 
So any discrepancies are likely the result of the $k_{\text{on}}$ numbers. Our hypotheses behind this discrepancy stem from the following realization. In the fluorescence quenching experiment \cite{Burkhard2000X-rayEnergies}, the binding affinity is estimated using Eq. \ref{eqn:quenching},
\begin{equation}
    \frac{\delta F}{\delta F_{\text{max}}} = \frac{[L]}{[L]+K_D},
    \label{eqn:quenching}
\end{equation}
where $\delta F$ is the change of Trp59 fluorescence due to ligand induced quenching and $[L]$ is the ligand concentration.
This equation is only valid when the ligand concentration is much higher than the protein concentration \cite{VanDeWeert2011FluorescenceMethodology}. At such high concentration, the FKBP protein will be surrounded by many ligands and the binding kinetics will be dominated by ligands situated close to the binding pocket. The effect of diffusion-dependent arrival from infinite distance may not be relevant in the context of the experiment. In the simulation work by Huang and Caflisch the trajectories were terminated after each ligand unbinding event, preventing the ligand to diffuse around in the solvent \cite{Huang2011TheUnbinding}. Consequently, their binding free energy results agree with the experimental values \cite{Burkhard2000X-rayEnergies} but differ from the results of extensive equilibrium MD \cite{Pan2017QuantitativeSimulations} and metadynamics simulations \cite{Pramanik2019CanDissociation} where ligand diffusion was taken into account. Our WEM and WEM-RR calculations also suffered from this problem, so the correction for the diffusion effect was necessary.

\begin{figure}[H]
\centering
\includegraphics[width=0.5\textwidth]{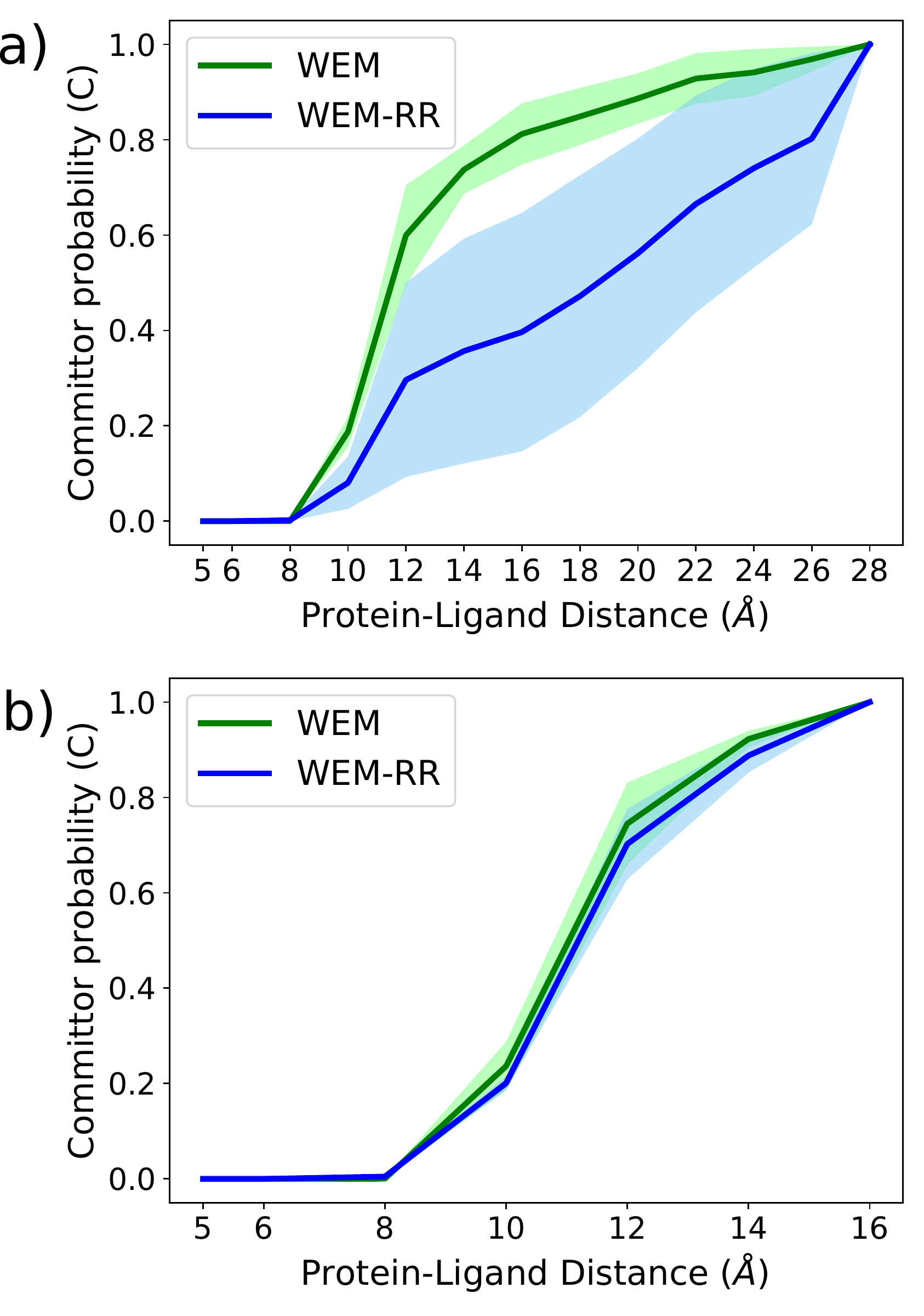}
\caption{Committor probabilities of BUT ligand dissociation from FKBP protein using traditional WEM and using WEM with harmonic restraint release. (a) unbound state is defined as the farthest milestone $r=28$\AA. (b) the milestone at $r=16$\AA$ $ is considered as the unbound state to avoid the diffusive regime. Error bars are standard deviation from 3 trial runs.
}
\label{fig:committor}
\end{figure}

The committor probability of ligand dissociation for trajectories starting at different milestones was calculated using the procedure described in section \ref{sec:WEMscheme}. The results for regular WEM and WEM-RR method (Fig. \ref{fig:committor}a) show very different absolute values but a common trend. The committor probability to reach the unbound state is near zero up to $r = 8$\AA, where it then shows a sharp increase until $r = 12$ \AA, after which it becomes linear. In the theoretical limit of a sharp $\delta$-function barrier the committor follows a step function dependence, switching its value from zero to one at the location of the barrier. Conversely a linearly increasing committor function is the signature of a flat free energy surface. The results in Figure \ref{fig:committor}a are consistent with the free energy profile in Fig. \ref{fig:WEM-freeE}a and \ref{fig:WEM-freeE}b. There is a sharp increase in $G$ at $6$\AA$<r<12$\AA$ $ followed by an essentially flat landscape. Because of the difference in the depth of the free energy minimum between WEM and WEM-RR calculation, the absolute values of the committor probability are different. As we identify $r=12$ \AA$ $ to be the beginning of diffusive regime, we chose $r=16$ \AA$ $ as our distance criterion for unbinding.

We also plotted the committors considering $r=16$\AA$ $ as the unbound state (Figure \ref{fig:committor}b), and where BUT is assumed to be transitioning to bulk solvent. In bulk solvent, BUT is subject to simple diffusion processes, where the free energy surface is relatively flat. The committor to $r=16$\AA$ $ therefore models the transition from protein-mediated dynamics to bulk-solvent dynamics. Now the committor distribution function of both WEM and WEM-RR methods agree with each other. Both indicate that the transition state for unbinding (with C = 0.5) should be between milestone $r=10$\AA$ $ and $r=12$\AA$ $. 

\begin{table}
\caption{\label{tab:Simulation_time}  Simulation time for WEM method  }
\begin{ruledtabular}
\begin{tabular}{cccc}

Method &Total simulation  &$r = 6$\AA$ $ &$r = 6$\AA$ $ molecular \\
 &time (ns) &milestone (ns) &time (ns) \\
\hline
\hline
WEM-RR trial 1 &56.64 &16.97 &0.54\\
WEM-RR trial 2 &63.80 &21.57 &0.69\\
WEM-RR trial 3 &83.15 &26.32 &0.80\\

\end{tabular}
\end{ruledtabular}

\end{table}

In table \ref{tab:Simulation_time} we showed the computational cost of the WEM simulation with harmonic restraint release. The values reported exclude the first 10 iterations for each starting point generation. They contributed about 10$\times$10$\times$2 ps = 200 ps, which is negligible compared to the total simulation time. Apparently converged results from WEM simulation can be obtained using a total simulation time of just $\sim$10$\langle \tau \rangle_{\text{off}}$ (Table \ref{tab:MFPT} \& \ref{tab:Simulation_time}). The convergence plots for the matrix elements of $\mathbf{K}$ are shown in Supplementary Materials.  On the average, only one or two binding/unbinding events could be observed from a continuous equilibrium MD simulation of that length for the same system \cite{Pan2017QuantitativeSimulations}. 

In table \ref{tab:Simulation_time} we also present the total simulation time of the bound state ($r = 6$\AA) as it requires the highest amount of total computational time to converge transition probabilities. A longer simulation time is often required for convergence for a milestone buried in deep free energy minimum as discussed in Ref. \citen{Ray2020WeightedSimulations}. 

Also, the individual molecular time (i.e., the time describing continuous phase space evolution as defined in Ref. \citen{Adhikari2019ComputationalTimes}) needed to converge the results for the bound state milestone is less than 1 ns (Table \ref{tab:Simulation_time}). It indicates that in presence of abundant computational resources, when we can parallelize each weighted ensemble trajectory for each milestone into a different core, we need to spend a wall clock time worth less than 1 ns to perform the entire WEM simulation. WEM, thus, provides significant reduction of computational cost without applying any biasing force. 

\begin{figure*}
\centering
\includegraphics[width=\textwidth]{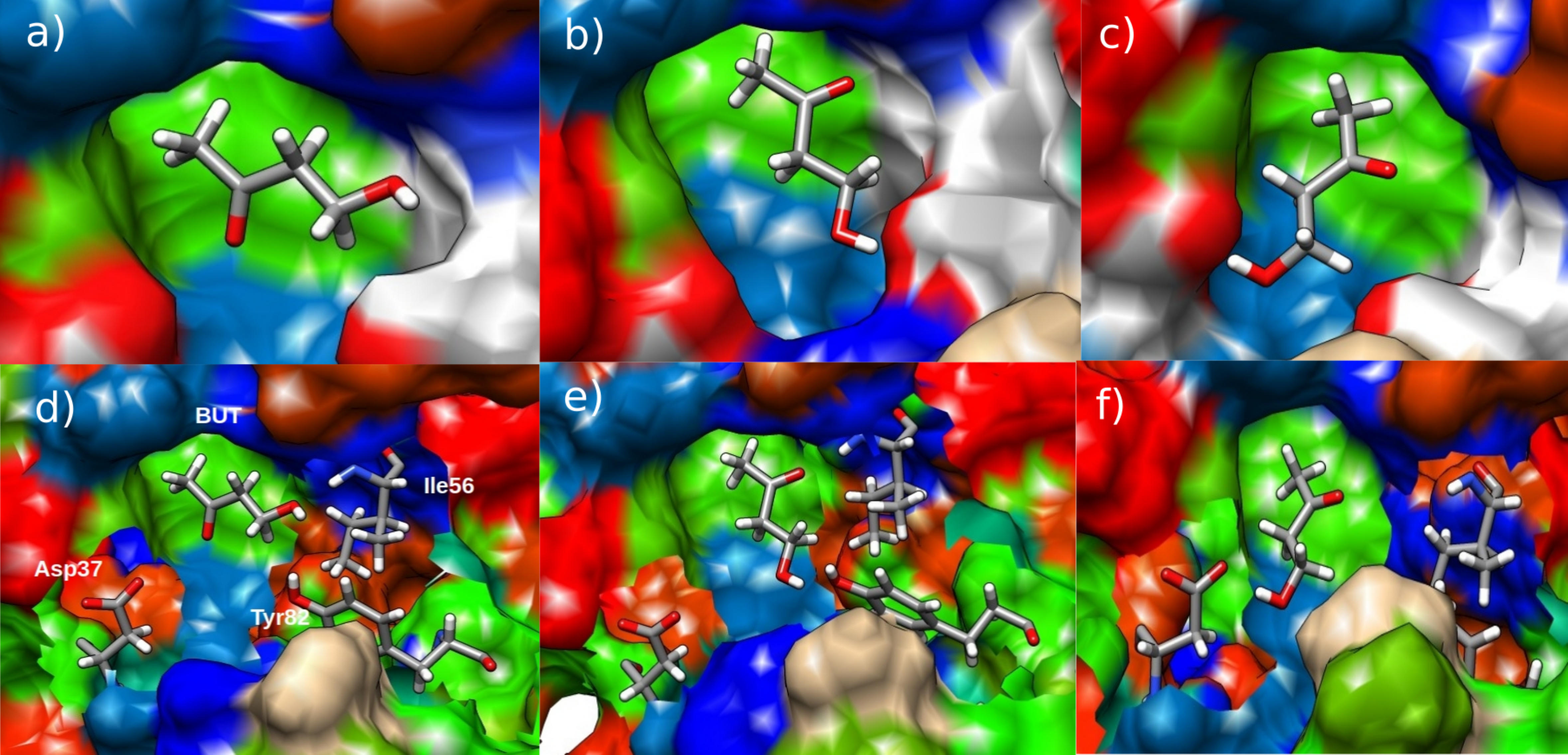}

\caption{Representative poses of BUT ligand in the binding pocket of FKBP protein; the binding pocket residues are color coded according to hydrophobicity  Upper panel (a, b, and c): the atoms of the ligand are explicitly shown in licorice. Lower panel (d, e, and f): The residues participating in hydrogen bonding with BUT are also shown explicitly in licorice}
\label{fig:poses}
\end{figure*}

\begin{figure*}
    \centering
    \includegraphics[scale=0.5]{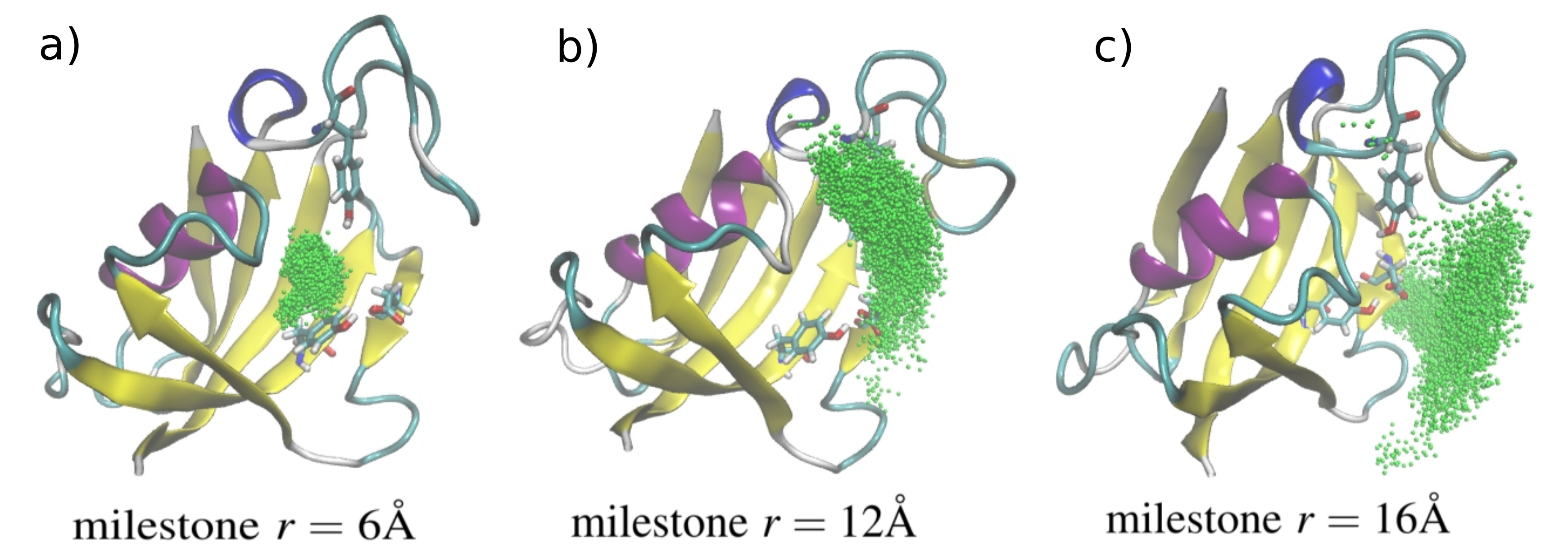}
    \caption{Distribution of the BUT ligand around the binding pocket of FKBP protein for three characteristic milestones ((a) bound state, (b) a milestone close to TS, and (c) the unbound state) using WEM-RR simulation. The method produces exhaustive sampling of pathways for the progression from bound to unbound states. In milestone $r = 6$\AA$ $ (a) the ligand is confined inside the binding pocket indicating that this milestone corresponds to the bound state. In milestone $r = 16${\AA} (c), which we assign to be the unbound state, the ligand is primary solvent exposed and far from the binding pocket. A possible transition state (TS) (which has committor = 0.5) is located in between milestone $r = 10$\AA$ $ and $r = 12$\AA$ $. The figure for milestone $r=12$\AA$ $ (b) shows that the ligand is not inside binding pocket but have significant interaction with the protein. For both the $r = 12$\AA$ $ and the unbound state, the ligand explores a wide range of conformations and not biased towards one ligand release pathway, although the starting structures for each milestone was sampled from one ligand release simulation. The conformations explored are in agreement with the results of Pramanik et al. \cite{Pramanik2019CanDissociation} and Dickson and Lotz \cite{Dickson2016LigandMechanisms}.}
    \label{fig:lig_dist}
\end{figure*}

The distribution of the ligand around the FKBP protein for different milestones has been depicted in Figure \ref{fig:lig_dist}. It indicates that the WEM simulation samples a large part of the configurational space around the binding site, potentially including multiple binding and unbinding pathways. Various ligand poses of BUT in the pocket and in the release site of the FKBP protein could also be sampled from WEM simulations pertaining to different milestones. 
 
We clustered the structures associated with milestone $r = 6$\AA$ $ (bound state) in the WEM-RR simulation using the root mean squared difference of inter-atomic distances to nearby residues computed with the UCSF Chimera package, \cite{Pettersen2004UCSFAnalysis} which implements the method proposed in Ref. \citen{LA1996AnSubfamilies}. The ligand poses are shown in Figure \ref{fig:poses}. BUT is observed to be interacting with Asp37 side-chain, Ile56 backbone and Tyr82 side-chain through hydrogen bonding. The ligand poses are in agreement with previous computational studies \cite{Huang2011TheUnbinding,Dickson2016LigandMechanisms}. This process can be repeated for other milestones near the binding pocket to search for additional meta-stable states.

\section{Concluding Discussion}
\label{sec:conclusions}
In summary, we applied and refined our combined weighted ensemble milestoning (WEM) scheme to study the binding and unbinding dynamics of protein-ligand systems. We studied a simple model of Na+/Cl- ion pair and a realistic model of BUT ligand bound to FKBP protein. The regular WEM method, proposed in our original publication \cite{Ray2020WeightedSimulations}, could calculate the residence time, binding and unbinding rate constants, binding affinity and free energy for the NaCl system with reasonable accuracy. However, the results for the FKBP-BUT system are far from accurate and call for serious modifications in our methodology. Particularly, the residence time is often 2-3 orders of magnitude different from the results obtained by conventional MD simulation and literature values, while the binding free energy is incorrect by several $k_BT$ units. The large variance of the results across multiple WEM runs is also a cause of concern. 

Here, we introduced a modification and extension of the WEM procedure to address these issues. Particularly, the regular WEM procedure was limited by the absence of multiple starting structures per milestone, and led to a strong dependence of the calculated values of the observables on the initial conditions. One possible solution could be to sample multiple structures from an umbrella sampling simulation with a progress variable confined to the hyperplane of the milestone and use them as starting points for multiple WE simulations. Indeed such an approach is followed in some of the recent milestoning-based host-guest binding studies \cite{Votapka2017SEEKR,Jagger2018QuantitativeApproach}. But in the weighted ensemble method a small number of trajectories are split into a very large number of trajectories with smaller weights. So this umbrella sampling based approach would increase the computational cost for WEM simulation to a large extent. \textcolor{black}{Previous studies showed that the majority ($\sim$80\%) of the total simulation time for milestoning simulations is spent in sampling initial structures on the milestone interface \cite{Ma2018ProbingMilestoning,Votapka2017SEEKR,Jagger2018QuantitativeApproach}. This was necessary because in regular milestoning the number of transition events sampled per milestone is, at most, equal to the number of starting trajectories. Because of the capability of WE method to spawn new trajectories on the fly, it is possible to obtain hundreds of transition events even from a single initial starting structure per milestone \cite{Ray2020WeightedSimulations}. However, as we realized that using only one starting point per milestone results in large variance in computed observables, we generated ten different structures for each milestone.} We achieved that by modifiying the WEM scheme by incorporating a harmonic restraint on the reaction coordinate for a few initial iterations of the weighted ensemble. It resulted in additional initial conditions from which unbiased WE simulation could be started. This approach, which we refer to as WEM restraint release (WEM-RR), involves a very simple workflow, only requiring the stopping and restarting of the WE simulation after the first few iterations and discarding the initial iterations from the analysis (so that the effect of the restraint is removed). However, the number of iterations required to generate the initial coordinates can vary from system to system, depending on the presence of slow orthogonal degrees of freedom. 

We showed that the WEM-RR method can be used to accurately calculate not only the residence time and binding free energy, but also the binding and unbinding rate constants and the binding affinity. The entire binding coordinate can be simultaneously sampled due to both the fine-grained parallelization of the WE procedure and the statistical independence of the milestones. \textcolor{black}{The binding free energy for the protein-ligand system is overestimated because of limited sampling of the milestone hypersurface (Figure \ref{fig:surface}) even when the ligand is far from the binding pocket. This ignores the possibility that the ligand can approach the protein or explore regions of 3D space at any solid angle when it is not bound. This effect is clear when we compare the PMF computed from WEM or WEM-RR simulations (Fig. \ref{fig:WEM-freeE}) to the one computed using metadynamics (see supplementary material). The metadynamics PMF decreases to a lower value at larger ligand-receptor distance because of the additional sampling of ligand configurations in the solvent. Consequently, while calculating the binding time, WEM runs were also unable to take into account the time spent by the ligand diffusing in solvent before arriving near the protein. This prompted the need to include this diffusion effect into our methodology. We proposed an approximate analytical method, which utilizes the WEM transition kernel, to treat the effect of diffusion in the calculation of binding time and rate constant. This technique could successfully reproduce the $k_{\text{on}}$ and $\Delta G^0$ values consistent with extensive MD simulation studies from literature and the enhanced sampling simulation independently performed herein.}

Our WEM procedure for the FKBP-BUT binding process required a net CPU time of less than 100 ns simulation. The fact that such a short simulation length produces kinetic results similar to microsecond simulations highlights the ability of the WEM procedure to rapidly evaluate a protein-ligand binding-unbinding process. With sufficient computational resources, the simulations for each milestone can be performed in parallel, providing significant gain in the wall clock time. \textcolor{black}{Yet, the comparison of simulation time with literature should be performed with caution. Except for the work by Pan et. al. \cite{Pan2017QuantitativeSimulations} all previous studies have calculated only one or two properties out a list containing the binding affinity, dissociation constant, and binding and unbinding kinetic constants. For example, Dickson and Lotz could compute the ligand residence time of the same system from 240 ns of WExplore simulation \cite{Dickson2016LigandMechanisms}. However, they could not report binding kinetics or binding free energy. One of the advantages of WEM (and milestoning based methods in general) is that the kinetics of \emph{both forward and backward} processes can be computed from one set of simulation data, leading to the availability of both the binding and unbinding rate constants, and the binding affinity. We therefore compared our simulation time only to the extensive equilibrium MD study by Pan et. al. }

\textcolor{black}{For both regular milestoning and  weighted ensemble based methods, the gain in computational efficiency is relatively higher for complex systems involving longer timescale ($\mu$s-ms), compared to systems with fast kinetics (ps-ns). Consequently, we believe the WEM method will also be highly efficient for simulating the binding-unbinding dynamics of more potent (low micromolar to nanomolar) ligands with longer residence time.} 

\textcolor{black}{This higher efficiency comes at the cost of complexity of the simulation protocol, involving an initial biased/unbiased MD to generate milestone anchors, performing WE simulations starting from those anchors and monitoring the trajectories for crossing of milestones. But due to the user friendly implementation of weighted ensemble method through the WESTPA \cite{Zwier2015} package, the overall complexity of the method is equivalent to regular milestoning simulation. We also share, with this paper, the codes of WEM and WEM-RR (see Data availability statement). The implementation is very generic, requiring only minor modifications for switching to a completely different system; and also includes scripts to automate most of the processes described in this work.}

\textcolor{black}{In the study of ligand-receptor dynamics, the choice of reaction coordinate (RC) is often a crucial factor. Our current methodology does not directly facilitate the choice of RC as it has to be picked beforehand. We used a single geometric distance based RC in our current study which could sufficiently reproduce the experimental observables from literature. However, for more complex systems, additional degrees of freedom, like solvation, protein conformational change, loop motion etc., can play an important role in ligand binding or unbinding; and their effect should be taken into consideration.} 

\textcolor{black}{Although it is theoretically possible to generalize the WEM protocol into higher dimension to include these collective variables, the number of milestones scales exponentially with the number of dimensions \cite{Elber2017CalculatingMilestoning}. So, it will lead to a significant increase in the complexity of the methodology, reducing its applicability in all practical purpose. We plan to address this issue in future work, by developing an easier protocol to include additional collective variables alongside the milestoning coordinate. Moreover, once one- or multi-dimensional milestones are defined, enhanced sampling protocols inside the milestones can be used to accelerate relaxation of slow degrees of freedom perpendicular to the reaction coordinate manifold.}

Other possible directions for further development of the current methodology are to perform simplified, Brownian dynamics (BD) simulation for distant milestones \cite{Votapka2017SEEKR,Jagger2018QuantitativeApproach}. This use of simplified dynamics for milestones far away from the binding site will further reduce the overall computational cost and increase the accuracy of $k_{\text{on}}$ and related properties. Importantly, this will also allow one to include the position dependence of the ligand diffusion constant using, e.g., a  Rotne-Prager diffusion tensor \cite{Skolnick2016Perspective:Macromolecules}. We plan to include this improvement into WEM in future work. 

Overall, the weighted ensemble milestoning (WEM) method can be used as a computationally inexpensive yet reliable tool to study kinetics and thermodynamics of molecular processes using all-atom MD simulation. The accuracy of the results are primarily limited by the choice of force field and the reaction coordinate, although an apparently simple reaction coordinate (the center of mass distance between protein and ligand) could provide results in agreement with experimental and simulation values. We believe that the WEM method, with its efficiency inherited from weighted ensemble, and its ability to calculate free energy and kinetics derived from milestoning, has the potential to become an essential tool for high level screening of protein inhibitors, and to play a key role in computational drug design.

\section*{Supplementary material}
See the supplementary material for the justification of milestone placing, details of metadynamics simulation, convergence plots of $K_D$, the sampled starting structures for WEM-RR simulation and the convergence plots of the elements of transition kernel for both the systems.

\section*{Acknowledgements}
The authors thank Alex Dickson for providing the BUT ligand parameters. DR thanks Ron Elber for a stimulating discussion and also for suggesting the idea to use velocity auto-correlation function to determine the appropriate spacing between milestones. IA acknowledges the support of the National Science Foundation (NSF) via grant MCB 2028443. DR was supported partially by the Molecular Science Software Institute seed fellowship funded by NSF via grant number OAC-1547580. The work has benefited from the computational resources of the UC Irvine High Performance Computing (HPC) cluster and San Diego Supercomputer Center (SDSC).

\section*{Disclosure}
The authors declare the following competing financial interest: DLM is a current member of the Scientific Advisory Board of OpenEye Scientific Software and an Open Science Fellow with Silicon Therapeutics.

\section*{Data availability statement}
The software codes of the WEM implementation, input files and analysis scripts specific to this work are available from the following github repository: \url{https://github.com/dhimanray/WEM-Ligand.git}. The code for calculating surface coverage using Monte Carlo method (see Appendix \ref{sec:diffusion}) is available from \url{https://github.com/trevorgokey/misc/tree/master/sphere_SA_population}. The data that supports the findings of this study are available within the article and the supplementary material. Additional data files can be made available from the authors upon request.

\appendix
\section{Diffusion effect on $k_{\text{on}}$}
\label{sec:diffusion}
The inward flux $k(r)$ of ligand at a spherical surface of radius $r$ is given by  \cite{McCammon1986DiffusionalAssociation}:
\begin{equation}
    k(r) = 4\pi D r
\end{equation}
\begin{figure}[H]
    \centering
    \includegraphics[width=0.5\textwidth]{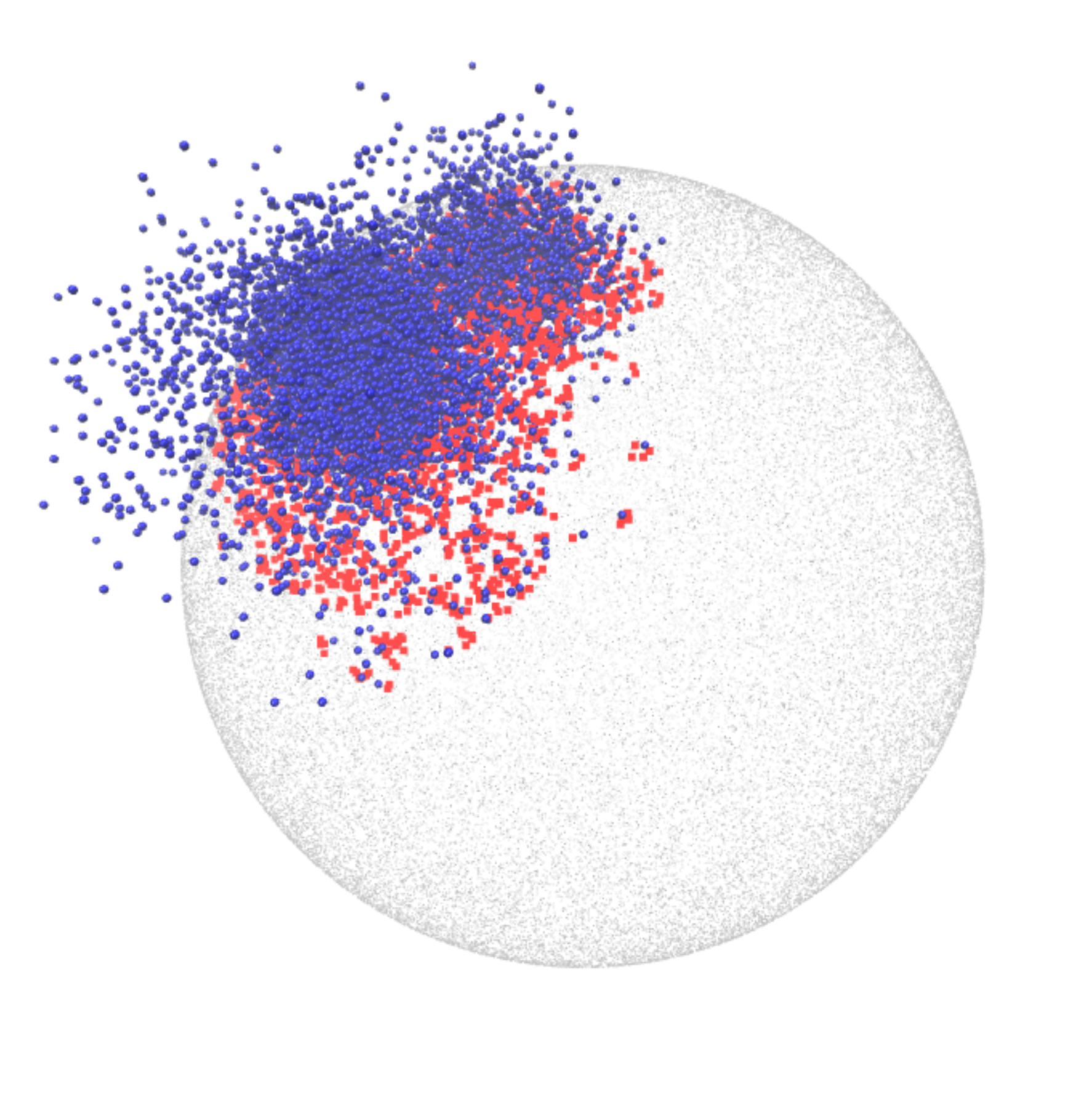}
	\caption{Computing surface coverage factor $\alpha$ in Eq. \ref{surface-coverage}  (see text). Surface area explored by the projection of BUT trajectories from milestone at $r=16$ \AA$ $; area calculated using 5 million points  uniformly sampled using Monte-Carlo method on the surface of the sphere. The ligand was represented as single point (in blue) and projected onto the milestone sphere. Randomly sampled points that had a great-circle distance between the arc connecting the ligand point and the sampled point, of less than 1 \AA$ $ contributed to the surface area covered. Sampled points that contributed are colored in red, whereas gray points did not contribute.  }
    \label{fig:surface}
\end{figure}
But trajectories arriving at all points of the sphere are not equally likely to propagate to the bound state. So we used a Monte Carlo based estimator to calculate the fraction of surface area of the sphere of radius $r$, explored by the ligand while being at milestone corresponding to radial distance $r$ (Figure \ref{fig:surface}). Let us say this fraction is $\alpha$. Some of the trajectories at that milestone will possibly propagate towards unbound state. The probability of traveling towards the binding pocket was estimated from the row-normalized milestoning transition matrix element $K_{i i-1}$. Including this two factors the flux of binding trajectories through milestone at distance $r$ is 
\begin{equation}
    k(r) = 4\pi D r \alpha K_{i i-1}
    \label{surface-coverage}
\end{equation}
In order to estimate the $k_{\text{on}}^{\text{diff}}$ due to diffusion in M$^{-1}$ s$^{-1}$ unit, considering the milestone at $r$ is the binding surface, we need to multiply with Avogadro's number $N_{av}$.
\begin{equation}
    k_{\text{on}}^{\text{diff}} = 4\pi D r \alpha K_{i i-1} N_{av}
\end{equation}
Hydrodynamics studies showed that at room temperature the diffusion constant of small organic molecules is around $1\times 10^5$ cm$^2$ s$^{-1}$ in water \cite{Delgado2007MolecularTemperatures}. Using this value for $D$ and after proper unit conversion the $k_{\text{on}}^{\text{diff}}$, in M$^{-1}$ s$^{-1}$ unit, is given by 
\begin{equation}
    k_{\text{on}}^{\text{diff}} = 7.569\times 10^8 r \alpha K_{i i-1} 
\end{equation}
where $r$ is measured in \AA. 

\newpage
\section*{References}
\bibliography{references}

\end{document}